%% file: main.tex
\documentclass[sigconf,screen]{acmart}

\AtBeginDocument{%
  \providecommand\BibTeX{{%
    \normalfont B\kern-0.5em{\scshape i\kern-0.25em b}\kern-0.8em\TeX}}}

\setcopyright{acmcopyright}
\acmPrice{15.00}
\acmDOI{10.1145/3293882.3330577}
\acmYear{2019}
\copyrightyear{2019}
\acmISBN{978-1-4503-6224-5/19/07}
\acmConference[ISSTA '19]{Proceedings of the 28th ACM SIGSOFT International Symposium  on  Software Testing and Analysis}{July 15--19, 2019}{Beijing, China}
\acmBooktitle{Proceedings of the 28th ACM SIGSOFT International Symposium  on  Software Testing and Analysis (ISSTA '19), July 15--19, 2019, Beijing, China}

\usepackage[ruled,vlined,lined,commentsnumbered]{algorithm2e}
\usepackage{booktabs}
\usepackage{moreverb}
\usepackage{fontenc}
\usepackage{amsmath}
\usepackage{amssymb}
\usepackage{fancybox}
\usepackage{color}
\usepackage{colortbl}
\usepackage{array}
\usepackage{multirow}
\usepackage{multicol}
\usepackage{listings}
\usepackage{makecell}
\usepackage{graphicx}
\usepackage{setspace}
\usepackage{footmisc}
\usepackage{wasysym}

\usepackage[ruled,vlined,lined,commentsnumbered]{algorithm2e}
\usepackage{balance}
\usepackage{csvsimple}
\usepackage{longtable}
\usepackage[skip=0pt,font=small,labelfont=bf]{caption}
\usepackage{url}
\usepackage{lscape}
\usepackage{rotating}
\usepackage{tikz}
\usepackage[normalem]{ulem}
\usepackage{enumitem}

\usepackage[most]{tcolorbox}

\usepackage{threeparttable}

\usepackage{marvosym}
\usepackage{pifont}

\newcolumntype{L}[1]{>{\raggedright\let\newline\\\arraybackslash\hspace{0pt}}m{#1}}
\newcolumntype{C}[1]{>{\centering\let\newline\\\arraybackslash\hspace{0pt}}m{#1}}
\newcolumntype{R}[1]{>{\raggedleft\let\newline\\\arraybackslash\hspace{0pt}}m{#1}}

\definecolor{codegreen}{rgb}{0,0.6,0}
\definecolor{codegray}{rgb}{0.5,0.5,0.5}
\definecolor{codepurple}{rgb}{0.58,0,0.82}
\definecolor{backcolour}{rgb}{0.95,0.95,0.92}
\definecolor{lightgray}{gray}{0.9}

\lstdefinestyle{mystyle}{
    commentstyle=\color{codegreen},
    keywordstyle=\color{magenta},
    numberstyle=\tiny\color{codegray},
    stringstyle=\color{codepurple},
    basicstyle=\footnotesize,
    breakatwhitespace=false,
    breaklines=true,
    captionpos=b,
    keepspaces=true,
    showspaces=false,
    showstringspaces=false,
    showtabs=false,
    tabsize=2
}

\lstdefinelanguage{diff}{
  morecomment=[f][\color{blue}]{@@},     
  morecomment=[f][\color{red}]-,         
  morecomment=[f][\color{codegreen}]+,       
  morecomment=[f][\color{red}]{---}, 
  morecomment=[f][\color{codegreen}]{+++},
}

\setlist{noitemsep} 

\definecolor{darkpastelred}{rgb}{0.76, 0.23, 0.13}
\definecolor{ao(english)}{rgb}{0.0, 0.5, 0.0}

\definecolor{darkpastelred}{rgb}{0.76, 0.23, 0.13}
\definecolor{ao(english)}{rgb}{0.0, 0.5, 0.0}

\lstdefinelanguage{diff}{
  morecomment=[f][\color{blue}]{@@},     
  morecomment=[f][\color{red}]-,         
  morecomment=[f][\color{codegreen}]+,       
  morecomment=[f][\color{red}]{---}, 
  morecomment=[f][\color{codegreen}]{+++},
}

\definecolor{yellow}{RGB}{255,255,153}
\definecolor{grey}{RGB}{224,224,224}

\newboolean{showcomments}
\setboolean{showcomments}{true}
\ifthenelse{\boolean{showcomments}}
 { \newcommand{\mynote}[2]{
      \fbox{\bfseries\sffamily\scriptsize#1}
        {\small$\blacktriangleright$\textsf{\emph{#2}}$\blacktriangleleft$}}}
        { \newcommand{\mynote}[2]{}}

\definecolor{DarkOrange}{rgb}{0.8,0.3,0.0}
\definecolor{DarkCyan}{rgb}{0.0, 0.55, 0.55}

\newcolumntype{?}{!{\vrule width 1pt}}

\newcommand{\nfixpatterns}{35\xspace}

\newcommand{\toolname}{\texttt{TBar}\xspace}

\newcommand{\fixpattern}[1]{
\vspace{-0.4cm}
\begin{tcolorbox}[tile,size=fbox,boxsep=-1mm,boxrule=0pt,top=0pt,bottom=0pt,
borderline west={2mm}{0pt}{black!5!white},colback=black!5!white] 
\em #1
\end{tcolorbox}
\vspace{-0.2cm}
}

\newcommand{\find}[1]{
\begin{tcolorbox}[tile,size=fbox,boxsep=0.3mm,boxrule=0pt,top=0pt,bottom=0pt,
borderline west={0.5mm}{0pt}{blue!5!white},colback=blue!5!white]
\em #1
\end{tcolorbox}
\vspace{-0.1cm}
}

\newcommand{\questions}[2]{
\begin{tcolorbox}[tile,size=fbox,boxsep=0.3mm,boxrule=0pt,top=0pt,bottom=0pt,title=#1,
borderline west={1mm}{0pt}{black!50!white},colback=black!5!white]
\em #2
\end{tcolorbox}
}

\begin{document}
\title{TBar: Revisiting {T}emplate-{B}ased {A}utomated Program {R}epair}

\author{Kui Liu, Anil Koyuncu, Dongsun Kim, Tegawend\'e F. Bissyand\'e}
\affiliation{%
  \institution{University of Luxembourg, Luxembourg}
}
\email{{kui.liu, anil.koyuncu, dongsun.kim, tegawende.bissyande}@uni.lu}

\input{abstract}

%
%



\maketitle

\input{Intro}

\input{background}

\input{fixpatterns}
\input{exp}
\input{empiricalStudy}
\input{discussion}
\input{relatedwork}
\input{conclusion}

\begin{acks}
This work is supported by the Fonds National de la Recherche (FNR), Luxembourg, through RECOMMEND 15/IS/10449467 and FIXPATTERN C15/IS/9964569.
\end{acks}

\balance
\bibliographystyle{ACM-Reference-Format}
\bibliography{bib/references}

\end{document}

%% file: abstract.tex
\begin{abstract}
We revisit the performance of template-based APR to
build comprehensive knowledge about the effectiveness of fix patterns, and
to highlight the importance of complementary steps such as fault localization or donor code retrieval.
To that end, we first investigate the literature to collect,
summarize and label recurrently-used fix patterns.
Based on the investigation, we build \toolname, a straightforward APR tool that
systematically attempts to apply these fix patterns
to program bugs. We thoroughly evaluate \toolname on the Defects4J benchmark.
In particular, we assess the actual qualitative and quantitative diversity of fix patterns,
as well as their effectiveness in yielding plausible or correct patches.
Eventually, we find that, assuming a perfect fault localization,
\toolname correctly/plausibly fixes 74/101 bugs. Replicating a standard and practical pipeline of APR assessment,
we demonstrate that \toolname correctly fixes 43 bugs from Defects4J,
an unprecedented performance in the literature (including all approaches,
i.e., template-based, stochastic mutation-based or synthesis-based APR).
\end{abstract}

%% file: Intro.tex
\section{Introduction}
\label{sec:intro}
Automated Program Repair (APR) has progressively become an essential research field. APR research is indeed promising to improve modern
software development by reducing the time and costs associated with program debugging tasks.
In particular, given that faults in software
cause substantial financial losses to the software
industry~\cite{nist2002Software,britton2013reversible},
there is a momentum in minimizing the time-to-fix intervals by APR.
Recently, various APR
approaches~\cite{nguyen2013semfix,westley2009automatically,le2012genprog,
kim2013automatic,coker2013program,ke2015repairing,mechtaev2015directfix,long2015staged,
le2016enhancing,le2016history,long2016automatic,chen2017contract,
le2017s3,long2017automatic,xuan2017nopol,xiong2017precise,jiang2018shaping,
wen2018context,hua2018towards,liu2019avatar,liu2019you,liu2019learning} have been proposed, aiming at reducing manual debugging efforts through
automatically generating patches.

An early strategy of APR is to generate concrete patches based
on fix patterns~\cite{kim2013automatic} (also referred
to as fix templates~\cite{liu2018mining} or program transformation schemas~\cite{hua2018towards}).
This strategy is now common in
the literature and has been implemented in several
APR systems~\cite{kim2013automatic,saha2017elixir,durieux2017dynamic,
liu2018mining,hua2018towards,koyuncu2018fixminer,martinez2018ultra,liu2019avatar,liu2019you}.
Kim et al.~\cite{kim2013automatic} showed the usefulness
of fix patterns with {\em PAR}.
Saha et al.~\cite{saha2017elixir} later proposed \emph{ELIXIR}
by adding three new patterns on top of {\em PAR}~\cite{kim2013automatic}.
Durieux et al.~\cite{durieux2017dynamic} proposed {\em NPEfix} to repair null pointer exception bugs, using nine pre-defined fix patterns.
Long et al. designed {\em Genesis}~\cite{long2017automatic} to infer fix patterns for
specific three classes of defects. 
Liu and Zhong~\cite{liu2018mining} explored posts from Stack Overflow
to mine fix patterns for APR.
Hua et al. proposed {\em SketchFix}~\cite{hua2018towards}, a runtime on-demand
APR tool with six pre-defined fix patterns.
Recently, Liu et al.~\cite{liu2019avatar} used the fix patterns of FindBugs static violations~\cite{liu2018mining2} to fix semantic bugs.
Concurrently, Ghanbari and Zhang~\cite{ghanbari2018practical} showed that straightforward application of fix patterns (i.e., mutators) on Java bytecode is effective for repair. They do not, however, provide a comprehensive assessment of the repair performance yielded by each implemented mutator.

Although the literature has reported promising results with fix patterns-based APR, to the best of our knowledge, no extensive assessment on the effectiveness of various patterns is performed. A few most recent approaches~\cite{liu2018mining,hua2018towards,liu2019avatar} reported which benchmark bugs are fixed by each of their patterns. Nevertheless, many relevant questions on the effectiveness of fix patterns remain unanswered.

{\bf This paper.} Our work thoroughly investigates to what extent fix patterns are effective for program repair. In particular, emphasizing on the recurrence of some patterns in APR, we dissect their actual contribution to repair performance.
Eventually, we explore three aspects of fix patterns:
\begin{itemize}[leftmargin=*]
	\item {\em Diversity}: How diverse are the fix patterns used by the state-of-the-art? We survey the literature to identify and summarize the available patterns with a clear taxonomy.
	\item {\em Repair performance}: How effective are the different patterns? In particular, we investigate the variety of real-world bugs that can be fixed, the dissection of repair results, 
and their tendency to yield plausible or correct patches.
	\item  {\em Sensitivity to fault localization noise}: Are all fix patterns similarly sensitive to the false positives yielded by fault localization tools? We investigate sensitivity by assessing plausible patches as well as the suspiciousness rank of correctly-fixed bug locations.
\end{itemize}

Towards realizing this study, we implement an automated patch generation system, \toolname (\underline{T}emplate-\underline{B}ased \underline{a}utomated program \underline{r}epair), with a super-set of fix patterns that are collected, summarized, curated and labeled from the literature data.
We evaluate \toolname on the Defects4J~\cite{just2014defects4j} benchmark, and provide the replication package in a public repository:
{\bf \url{https://github.com/SerVal-DTF/TBar}}.
%

Overall, our investigations have yielded the following findings:
\begin{enumerate}[leftmargin=*]
	\item {\bf Record performance:} \toolname creates a new higher baseline of repair performance:
	74/101 bugs are correctly/plausibly fixed with perfect fault localization information and
	43/81 bugs are fixed with realistic fault localization output, respectively.
	\item {\bf Fix pattern selection:}
	Most bugs are correctly fixed only by a single fix pattern while other patterns generate
	plausible patches. This implies that appropriate pattern prioritization can prevent
	from plausible/incorrect patches. Otherwise, APR tools might be overfitted in plausible but incorrect patches.
	\item {\bf Fix ingredient retrieval:}
	It is challenging for template-based APR to select appropriate
	donor code, which is an ingredient of patch generation when using fix patterns.
	Inappropriate donor code may cause plausible but incorrect patch generation.
	This motivates a new research direction: \emph{donor code prioritization}.
	\item {\bf Fault localization noise:}
	It turns out that fault localization accuracy has a large impact on
	repair performance when using fix patterns in APR (e.g., applying
	a fix pattern to incorrect location yields plausible/incorrect patches).
\end{enumerate}


%% file: background.tex
%

%% file: fixpatterns.tex
\section{Fix Patterns}
\label{sec:fp}

For this study, we  systematically review\footnote{For conferences and journals, we consider ICSE, FSE, ASE, ISSTA, ICSME, SANER, TSE, TOSEM, and EMSE. The search keywords are `program'+`repair', `bug' +`fix'.} the APR literature to identify approaches that leverage fix patterns.
Concretely, we consider the program repair website~\cite{programRepair}, a bibliography survey of APR~\cite{monperrus2018automatic}, proceedings of software engineering conference venues and journals as the source of relevant literature.
We focus on approaches dealing with Java program bugs, and manually collect, from the paper descriptions as well as the associated artifacts, all pattern instances that are explicitly mentioned.
Table~\ref{tab:FPPapers} summarizes the identified relevant literature and the quantity of identified fix patterns targeting Java programs. Note that the techniques described in the last four papers (i.e., HDRepair, ssFix, CapGen, and SimFix papers) do not directly use fix patterns: they leverage code change operators or rules, which we consider similar to using fix patterns.

\input{tables/fpworks.tex}

\subsection{Fix Patterns Inference}
Fix patterns have been explored with the following four ways:
\begin{enumerate}[leftmargin=*]
	\item {\bf Manual Summarization}: Pan et al.~\cite{pan2009toward} identified 27 fix patterns from
	 	patches of five Java projects to characterize the fix ingredients of patches. They do not however apply the identified patterns to fix actual bugs.
		Motivated by this work, Kim et al.~\cite{kim2013automatic} summarized 10 fix patterns manually extracted from 62,656 human-written patches collected from Eclipse JDT.
	\item {\bf Mining}: Long et al.~\cite{long2017automatic} proposed Genesis, to infer fix patterns for three kinds of defects from existing patches. Liu and
	Zhong~\cite{liu2018mining} explored fix patterns from Q\&A posts in Stack
	Overflow. Koyuncu et al.~\cite{koyuncu2018fixminer} mined fix patterns at the AST level from patches by using code change differentiating tool~\cite{falleri2014fine}.
	Liu et al.~\cite{liu2018mining2} and Rolim et al.~\cite{rolim2018learning}
	proposed to mine fix patterns for static analysis violations.
	In general, mining approaches yield a large number of fix patterns, which are not always about addressing deviations in program behavior. For example, many patterns are about code style~\cite{liu2019avatar}. Recently, with {\em AVATAR}~\cite{liu2019avatar}, we proposed an APR tool that considers static analysis violation fix patterns to fix semantic bugs. 
	\item {\bf Pre-definition}: Durieux et al.~\cite{durieux2017dynamic} pre-defined 9 repair
	actions for null pointer exceptions by unifying the related fix patterns proposed in previous studies~\cite{dobolyi2008changing,kent2008dynamic,long2014automatic}. On the top of {\em PAR}~\cite{kim2013automatic}, Saha et al.~\cite{saha2017elixir} further defined 3 new fix
	patterns to improve the repair performance. Hua et al.~\cite{hua2018towards} proposed an
	APR tool with six pre-defined so-called code transformation schemas.
	We also consider operator mutations~\cite{martinez2016astor} as  pre-defined fix patterns, as the number of operators and mutation possibilities is limited and pre-set.
	Xin and Reiss~\cite{xin2017leveraging} proposed an approach to fixing bugs with 34 predefined code change rules at the AST level.
	Ten of the rules are not for transforming the buggy code but for the simple replacement
	of multi-statement code fragments. We discard these rules from our study to limit bias.
	\item {\bf Statistics}: Besides formatted fix patterns, researchers~\cite{wen2018context,jiang2018shaping} also explored
	to automate program repair with code change instructions (at the abstract syntax tree level) that are statistically recurrent in existing patches~\cite{zhong2015empirical,martinez2015mining, liu2018closer,wen2017empirical, jiang2018shaping}. The strategy is then to select the top-{\em n} most frequent code change instructions as fix ingredients to synthesize patches.
\end{enumerate}

\subsection{Fix Patterns Taxonomy}
After manually assessing all fix patterns presented in the literature (cf. Table~\ref{tab:FPPapers}), we identified
15 categories of patterns labeled based on the code context (e.g., {\em a cast expression}), the code change actions (e.g., {\em insert an ``if'' statement with ``instanceof'' check}) as well as the targets (e.g., {\em ensure the program will no throw a ClassCastException.}).
A given category may include one or several specialized sub-categories. Below, we present the labeled categories and provide the associated 35 {\bf Code Change Patterns} described in simplified GNU diff pattern for easy understanding.

\noindent
\underline{\bf FP1. Insert Cast Checker.} Inserting an {\em instanceof} check before one buggy statement if this statement contains at least one unchecked cast expression.
{\bf Implemented in:} PAR, Genesis, AVATAR, SOFix$^\dagger$, HDRepair$^\dagger$, SketchFix$^\dagger$, CapGen$^\dagger$, and SimFix$^\dagger$.\\
	\fixpattern{
\begin{lstlisting}[language=java,linewidth={\linewidth},basicstyle=\scriptsize\ttfamily]
(@\phantom{x}\hspace{6.7ex} \textcolor{codegreen}{+}@)  if (exp instanceof T) {
(@\phantom{x}\hspace{14ex}@) var = (T) exp; ......
(@\phantom{x}\hspace{6.7ex} \textcolor{codegreen}{+}@)  }
\end{lstlisting}}
	\noindent
	where {\em exp} is an expression (e.g., a variable expression) and {\em T} is the casting type, while ``$\ldots\ldots$'' means the subsequent statements dependent on the variable {\em var}. 
	Note that, ``$\dagger$'' denotes that the fix pattern is not specifically illustrated in the corresponding APR tools since the tools have some abstract fix patterns that can cover the fix pattern. The same notation applies to the following descriptions.

\noindent
\underline{\bf FP2. Insert Null Pointer Checker.} Inserting a {\em null} check before a buggy statement if, in this statement, a field or an expression (of non-primitive data type) is accessed without a null pointer check.
{\bf Implemented in:} PAR, ELIXIR, NPEfix, Genesis, FixMiner, AVATAR, HDRepair$^\dagger$, SOFix$^\dagger$, SketchFix$^\dagger$, CapGen$^\dagger$, and SimFix$^\dagger$.\\
    \fixpattern{
\begin{lstlisting}[language=java,linewidth={\linewidth},basicstyle=\scriptsize\ttfamily]
(@\phantom{x}\hspace{0ex}{\bf FP2.1:}\hspace{2ex} \textcolor{codegreen}{+}@)  if (exp != null) {   
(@\phantom{x}\hspace{15ex}@) ...exp...; ......
(@\phantom{x}\hspace{7.7ex} \textcolor{codegreen}{+}@)  }
(@\phantom{x}{\bf FP2.2:}\hspace{2ex} \textcolor{codegreen}{+}@)  if (exp == null) return DEFAULT_VALUE;  
(@\phantom{x}\hspace{11ex}@) ...exp...;
(@\phantom{x}{\bf FP2.3:}\hspace{2ex} \textcolor{codegreen}{+}@)  if (exp == null) exp = exp1;  
(@\phantom{x}\hspace{11ex}@) ...exp...;
(@\phantom{x}{\bf FP2.4:}\hspace{2ex} \textcolor{codegreen}{+}@)  if (exp == null) continue;  
(@\phantom{x}\hspace{11ex}@) ...exp...;
(@\phantom{x}{\bf FP2.5:}\hspace{2ex} \textcolor{codegreen}{+}@)  if (exp == null)
(@\phantom{x}\hspace{7.8ex} \textcolor{codegreen}{+}\hspace{6ex}@)throw new IllegalArgumentException(...);  
(@\phantom{x}\hspace{11ex}@) ...exp...;
\end{lstlisting}}
\noindent
where {\tt DEFAULT\_VALUE} is set based on the return type (RT) of the encompassing method as below:
\begin{equation}
\scriptsize
	\begin{array}{l}
    	{\tt DEFAULT\_{VALUE}} =
    	\begin{cases}
			\text{false}, &  \text{if } RT = boolean; \\
			0, &  \text{if } RT = {primitive}\text{  }{type}; \\
			\text{new String()}, &  \text{if } RT = String; \\
			\text{``return;''}, &  \text{if } RT = void; \\
 			\text{null}, & otherwise.
 		\end{cases}
 	\end{array}
\label{eq:normalization}
\end{equation}
{\em exp1} is a compatible expression in the buggy program (i.e., that has the same data type as {\em exp}).
{\bf FP2.4} is specific to the case of a buggy statement within a loop (i.e., {\em for} or {\em while}).

\noindent
\underline{\bf FP3. Insert Range Checker.}
Inserting a range checker for the access of an array or collection if it is unchecked.
{\bf Implemented in:} PAR, ELIXIR, Genesis, SketchFix, AVATAR, SOFix$^\dagger$ and SimFix$^\dagger$.\\
\fixpattern{
\begin{lstlisting}[language=java,linewidth={\linewidth},basicstyle=\scriptsize\ttfamily]
(@\phantom{x}\hspace{6.7ex} \textcolor{codegreen}{+}@)  if (index < exp.length) {
(@\phantom{x}\hspace{14ex}@) ...exp[index]...; ......
(@\phantom{x}\hspace{6.7ex} \textcolor{codegreen}{+}@)  }
(@\phantom{xx}@)OR
(@\phantom{x}\hspace{6.7ex} \textcolor{codegreen}{+}@)  if (index < exp.size()) {
(@\phantom{x}\hspace{14ex}@) ...exp.get(index)...; ......
(@\phantom{x}\hspace{6.7ex} \textcolor{codegreen}{+}@)  }

\end{lstlisting}}
\noindent
	where {\em exp} is an expression representing an array or collection.

\noindent
\underline{\bf FP4. Insert Missed Statement.} Inserting a missing statement before, or after, or surround a buggy statement. The statement is either an expression statement with a method invocation, or a {\em return/try-catch/if} statement.
{\bf Implemented in:} ELIXIR, HDRepair, SOFix, SketchFix, CapGen, FixMiner, and SimFix.\\
	\fixpattern{
\begin{lstlisting}[language=java,linewidth={\linewidth},basicstyle=\scriptsize\ttfamily]
(@\phantom{x}{\bf FP4.1:}\hspace{2ex} \textcolor{codegreen}{+}@)  method(exp);
(@\phantom{x}{\bf FP4.2:}\hspace{2ex} \textcolor{codegreen}{+}@)  return DEFAULT_VALUE;
(@\phantom{x}{\bf FP4.3:}\hspace{2ex} \textcolor{codegreen}{+}@)  try {
(@\phantom{x}\hspace{15ex}@) statement; ......
(@\phantom{x}\hspace{7.7ex} \textcolor{codegreen}{+}@)  } catch (Exception e) {   ...   }
(@\phantom{x}{\bf FP4.4:}\hspace{2ex} \textcolor{codegreen}{+}@)  if (conditional_exp) {
(@\phantom{x}\hspace{15ex}@) statement; ......
(@\phantom{x}\hspace{7.7ex} \textcolor{codegreen}{+}@)  }
\end{lstlisting}}
\noindent
	where {\em exp} is an expression from a buggy {\em statement}. It may be empty if the method does not take any argument. {\bf FP4.4} excludes three fix patterns ({\bf FP1}, {\bf FP2}, and {\bf FP3}) that are used with specific contexts.

\noindent
\underline{\bf FP5. Mutate Class Instance Creation.} Replacing a class instance creation expression with a cast {\em super.clone()} method invocation if the class instance creation is in an overridden clone method.
{\bf Implemented in:} AVATAR.\\
	\fixpattern{
\begin{lstlisting}[language=java,linewidth={\linewidth},basicstyle=\scriptsize\ttfamily]
(@\phantom{x}\hspace{6.7ex}@) public Object clone() {
(@\phantom{x}\hspace{6.7ex} \textcolor{red}{-}\hspace{4ex}@)... new T();
(@\phantom{x}\hspace{6.7ex} \textcolor{codegreen}{+}\hspace{4ex}@)... (T) super.clone();
(@\phantom{x}\hspace{6.7ex}@) }
\end{lstlisting}}
\noindent
	where {\em T} is the class name of the current class containing the buggy statement.

\noindent
\underline{\bf FP6. Mutate Conditional Expression.} Mutating a conditional expression that returns a boolean value (i.e., {\tt true} or {\tt false}) by either updating it, or removing a sub conditional expression, or inserting a new conditional expression into it.
{\bf Implemented in:} PAR, ssFix, S3, HDRepair, ELIXIR, SketchFix, CapGen, SimFix, and AVATAR.\\
	\fixpattern{
\begin{lstlisting}[language=java,linewidth={\linewidth},basicstyle=\scriptsize\ttfamily]
(@\phantom{x}{\bf FP6.1:}\hspace{2ex} \textcolor{red}{-}@)  ...condExp1...
(@\phantom{x}\hspace{7.8ex} \textcolor{codegreen}{+}@)  ...condExp2...
(@\phantom{x}{\bf FP6.2:}\hspace{2ex} \textcolor{red}{-}@)  ...condExp1 Op condExp2...
(@\phantom{x}\hspace{7.8ex} \textcolor{codegreen}{+}@)  ...condExp1...
(@\phantom{x}{\bf FP6.3:}\hspace{2ex} \textcolor{red}{-}@)  ...condExp1...
(@\phantom{x}\hspace{7.8ex} \textcolor{codegreen}{+}@)  ...condExp1 Op condExp2...
\end{lstlisting}}
	\noindent
	where {\em condExp1} and {\em condExp2} are conditional expressions. {\em Op} is the logical operator `||' or `\&\&'. The mutation of operators in conditional expressions is not summarized in this fix pattern but in {\bf FP11}.

\noindent
\underline{\bf FP7. Mutate Data Type.} Replacing the data type in a variable declaration or a cast expression with another data type.
{\bf Implemented in:} PAR, ELIXIR, FixMiner, SOFix, CapGen, SimFix, AVATAR, and HDRepair$^\dagger$.\\
	\fixpattern{
\begin{lstlisting}[language=java,linewidth={\linewidth},basicstyle=\scriptsize\ttfamily]
(@\phantom{x}{\bf FP7.1:}\hspace{2ex} \textcolor{red}{-}@)  T1 var ...;
(@\phantom{x}\hspace{7.8ex} \textcolor{codegreen}{+}@)  T2 var ...;
(@\phantom{x}{\bf FP7.2:}\hspace{2ex} \textcolor{red}{-}@)  ...(T1) exp...;
(@\phantom{x}\hspace{7.8ex} \textcolor{codegreen}{+}@)  ...(T2) exp...;
\end{lstlisting}}
\noindent
	where both {\em T1} and {\em T2} denote two different data types. {\em exp} means the being casted expression (including variable).

\noindent
\underline{\bf FP8. Mutate Integer Division Operation.} Mutating the integer division expressions to return a float value, by mutating its divisor or divider to make them be of type float.
{\bf Released by} Liu et al.~\cite{liu2018mining2}, it is not implemented in any APR tool yet. \\
	\fixpattern{
\begin{lstlisting}[language=java,linewidth={\linewidth},basicstyle=\scriptsize\ttfamily]
(@\phantom{x}{\bf FP8.1:}\hspace{2ex} \textcolor{red}{-}@)  ...dividend / divisor...
(@\phantom{x}\hspace{7.8ex} \textcolor{codegreen}{+}@)  ...dividend / (double or float) divisor...
(@\phantom{x}{\bf FP8.2:}\hspace{2ex} \textcolor{red}{-}@)  ...dividend / divisor...
(@\phantom{x}\hspace{7.8ex} \textcolor{codegreen}{+}@)  ...(double or float) dividend / divisor...
(@\phantom{x}{\bf FP8.3:}\hspace{2ex} \textcolor{red}{-}@)  ...dividend / divisor...
(@\phantom{x}\hspace{7.8ex} \textcolor{codegreen}{+}@)  ...(1.0 / divisor) * dividend...
\end{lstlisting}}
\noindent
	where {\em dividend} and {\em divisor} are integer number literals or integer-returned expressions (including variables).

\noindent
\underline{\bf FP9. Mutate Literal Expression.} Mutating  boolean, number, or String literals in a buggy statement with other relevant literals, or correspondingly-typed expressions.
{\bf Implemented in:} HDRepair, S3, FixMiner, SketchFix, CapGen, SimFix and ssFix$^\dagger$.\\
	\fixpattern{
\begin{lstlisting}[language=java,linewidth={\linewidth},basicstyle=\scriptsize\ttfamily]
(@\phantom{x}{\bf FP9.1:}\hspace{2ex} \textcolor{red}{-}@)  ...literal1...
(@\phantom{x}\hspace{7.8ex} \textcolor{codegreen}{+}@)  ...literal2...
(@\phantom{x}{\bf FP9.2:}\hspace{2ex} \textcolor{red}{-}@)  ...literal1...
(@\phantom{x}\hspace{7.8ex} \textcolor{codegreen}{+}@)  ...exp...
\end{lstlisting}}
\noindent
	where {\em literal1} and {\em literal2} are of the same type literals, but having different values (e.g., {\em literal1} is {\tt true}, {\em literal2} is {\tt false}). {\em exp} denotes any expression value of the same type as {\em literal1}.

\noindent
\underline{\bf FP10. Mutate Method Invocation Expression.} Mutating the bu-ggy method invocation expression by adapting its method name or arguments. This pattern consists of four sub fix patterns:
\begin{enumerate}
	\item Replacing the method name with another one which has a compatible return type and same parameter type(s) as the buggy method that was invoked.
	\item Replacing at least one argument with another expression which has a compatible data type. Replacing a literal or variable is not included in this fix pattern, but rather in {\bf FP9} and {\bf FP13} respectively.
	\item Removing argument(s) if the method invocation has the suitable overridden methods.
	\item Inserting argument(s) if the method invocation has the suitable overridden methods.
\end{enumerate}
{\bf Implemented in:} PAR, HDRepair, ssFix, ELIXIR, FixMiner, SOFix, SketchFix, CapGen, and SimFix.\\
	\fixpattern{
\begin{lstlisting}[language=java,linewidth={\linewidth},basicstyle=\scriptsize\ttfamily]
(@\phantom{x}{\bf FP10.1:}\hspace{2ex} \textcolor{red}{-}@)  ...method1(args)...
(@\phantom{x}\hspace{8.8ex} \textcolor{codegreen}{+}@)  ...method2(args)...
(@\phantom{x}{\bf FP10.2:}\hspace{2ex} \textcolor{red}{-}@)  ...method1(arg1, arg2, ...)...
(@\phantom{x}\hspace{8.8ex} \textcolor{codegreen}{+}@)  ...method1(arg1, arg3, ...)...
(@\phantom{x}{\bf FP10.3:}\hspace{2ex} \textcolor{red}{-}@)  ...method1(arg1, arg2, ...)...
(@\phantom{x}\hspace{8.8ex} \textcolor{codegreen}{+}@) ... method1(arg1, ...)...
(@\phantom{x}{\bf FP10.4:}\hspace{2ex} \textcolor{red}{-}@)  ...method1(arg1, ...)...
(@\phantom{x}\hspace{8.8ex} \textcolor{codegreen}{+}@)  ...method1(arg1, arg2, ...)...
\end{lstlisting}}
\noindent
	where {\em method1} and {\em method2} are the names of invoked methods. {\em args}, {\em arg1}, {\em arg2} and {\em arg3} denote the argument expressions in the method invocation. Note that, code changes on class instance creation, constructor and super constructor expressions are also included in these four fix patterns.

\noindent
\underline{\bf FP11. Mutate Operators.} Mutating an operation expression by mutating its operator(s). We divide this fix pattern into three sub-fix patterns following the operator types and mutation actions.
\begin{enumerate}
	\item Replacing one operator with another operator from the same operator class (e.g., relational or arithmetic).
	\item Changing the priority of arithmetic operators.
	\item Replacing {\tt instanceof} operator with (in)equality operators.
\end{enumerate}
{\bf Implemented in:} HDRepair, ssFix, ELIXIR, S3, jMutRepair, SOFix, FixMiner, SketchFix, CapGen, SimFix, AVATAR, and PAR$^\dagger$.\\
	\fixpattern{
\begin{lstlisting}[language=java,linewidth={\linewidth},basicstyle=\scriptsize\ttfamily]
(@\phantom{x}{\bf FP11.1:}\hspace{2ex} \textcolor{red}{-}@)  ...exp1 Op1 exp2...
(@\phantom{x}\hspace{8.8ex} \textcolor{codegreen}{+}@)  ...exp1 Op2 exp2...
(@\phantom{x}{\bf FP11.2:}\hspace{2ex} \textcolor{red}{-}@)  ...(exp1 Op1 exp2) Op2 exp3...
(@\phantom{x}\hspace{8.8ex} \textcolor{codegreen}{+}@)  ...exp1 Op1 (exp2 Op2 exp3)... 
(@\phantom{x}{\bf FP11.3:}\hspace{2ex} \textcolor{red}{-}@)  ...exp instanceof T...
(@\phantom{x}\hspace{8.8ex} \textcolor{codegreen}{+}@)  ...exp != null...
\end{lstlisting}}
\noindent
	where {\em exp} denotes the expressions in the operation  and {\em Op} is the associated operator. 

\noindent
\underline{\bf FP12. Mutate Return Statement.} Replacing the expression (excluding literals, variables, and conditional expressions) in a return statement with a compatible expression.
{\bf Implemented in:} ELIXIR, SketchFix, and HDRepair$^\dagger$.\\
	\fixpattern{
\begin{lstlisting}[language=java,linewidth={\linewidth},basicstyle=\scriptsize\ttfamily]
(@\phantom{x}\hspace{6.7ex} \textcolor{red}{-}@)   return exp1;
(@\phantom{x}\hspace{6.7ex} \textcolor{codegreen}{+}@)   return exp2;
\end{lstlisting}}
\noindent
where {\em exp1} and {\em exp2} represent the returned expressions.

\noindent
\underline{\bf FP13. Mutate Variable.} Replacing a variable in a buggy statement with a compatible expression (including variables and literals).
{\bf Implemented in:} S3, SOFix, FixMiner, SketchFix, CapGen, SimFix, AVATAR, and ssFix$^\dagger$.\\
	\fixpattern{
\begin{lstlisting}[language=java,linewidth={\linewidth},basicstyle=\scriptsize\ttfamily]
(@\phantom{x}{\bf FP13.1:}\hspace{2ex} \textcolor{red}{-}@)  ...var1...
(@\phantom{x}\hspace{8.8ex} \textcolor{codegreen}{+}@)  ...var2...
(@\phantom{x}{\bf FP13.2:}\hspace{2ex} \textcolor{red}{-}@)  ...var1...
(@\phantom{x}\hspace{8.8ex} \textcolor{codegreen}{+}@)  ...exp...
\end{lstlisting}}
\noindent
where {\em var1} denotes a variable in the buggy statement. {\em var2} and {\em exp} represent respectively a compatible variable and expression of the same type as {\em var1}.

\noindent
\underline{\bf FP14. Move Statement.} Moving a buggy statement to a new position.
{\bf Implemented in:} PAR.\\
	\fixpattern{
\begin{lstlisting}[language=java,linewidth={\linewidth},basicstyle=\scriptsize\ttfamily]
(@\phantom{x}\hspace{7.7ex} \textcolor{red}{-}\hspace{4ex}@)statement;
(@\phantom{x}\hspace{12.7ex}@) ......
(@\phantom{x}\hspace{7.7ex} \textcolor{codegreen}{+}\hspace{4ex}@)statement;
\end{lstlisting}}
\noindent
where {\em statement} represents the buggy statement.

\noindent
\underline{\bf FP15. Remove Buggy Statement.} Deleting entirely the buggy statement from the program.
{\bf Implemented in:} HDRepair, SOFix, FixMiner, CapGen, and AVATAR.\\
	\fixpattern{
\begin{lstlisting}[language=java,linewidth={\linewidth},basicstyle=\scriptsize\ttfamily]
(@\phantom{x}{\bf FP15.1:}\hspace{7ex}@)......
(@\phantom{x}\hspace{7.8ex} \textcolor{red}{-}\hspace{4ex}@)statement;
(@\phantom{x}\hspace{11ex}@) ......
(@\phantom{x}{\bf FP15.2:}\hspace{2ex}\textcolor{red}{-}@)  methodDeclaration(Arguments) {
(@\phantom{x}\hspace{7.8ex} \textcolor{red}{-}\hspace{4ex}@)......; statement;......
(@\phantom{x}\hspace{7.8ex} \textcolor{red}{-}@)  }
\end{lstlisting}}
\noindent
where {\em statement} denotes any identified buggy statement, and {\em method} represents the encompassing method.

\subsection{Analysis of Collected Patterns}
We provide a study of the collected fix patterns following quantitative (overall set) and qualitative (per fix pattern) aspects.
Table~\ref{tab:fpQualitative} assesses the fix patterns in terms of four qualitative dimensions:
\begin{enumerate}[leftmargin=*]
	\item {\bf Change Action}: what high-level operations are applied on a buggy code entity? On the one hand, {\em Update} operations replace the buggy code entity with retreived donor code, while {\em Delete} operations just remove the buggy code entity from the program. On the other hand, {\em Insert} operations insert an otherwise missing code entity into the program, and {\em Move} operations change the position of the buggy code entity to a more suitable location in the program.
	\item {\bf Change Granularity}: what kinds of code entities are directly impacted by the change actions? This entity can be an entire {\em Method}, a whole {\em Statement} or specifically targeting an {\em Expression} within a statement.
	\item {\bf Bug Context}: what specific AST nodes of code entities are used to match fix patterns.
	\item {\bf Change Spread}: the number of statements impacted by each fix pattern.
\end{enumerate}

Quantitatively, as summarized in Table~\ref{tab:fpActions}, 17 fix patterns are related to {\em Update} change actions, 4 fix patterns implement {\em Delete} actions, 13 fix patterns {\em Insert} extra code, and only 1 fix pattern is associated to {\em Move} change action. 

In terms of change granularity, 21 and 17 fix patterns are applied respectively at the expression and statement code entity levels \footnote{Among these, four sub-fix patterns ({\bf FP10}) can be applied to either expressions or statements, given that constructor and super-constructor code entities in Java program are grouped into statement level in terms of abstract syntax tree by Eclipse JDT.}. Only 1 fix pattern is suitable at the method level.

Overall, we note that 30 fix patterns are applicable to a single  statement, while 7 fix patterns can mutate multiple statements at the same time. Among these patterns, {\bf FP14} and {\bf FP15.1} can both mutate single and multiple statements.

\input{tables/fpQualitative.tex}

\input{tables/fpActions.tex}

\begin{figure*}[!t]
    \centering
    \includegraphics[width=0.85\linewidth]{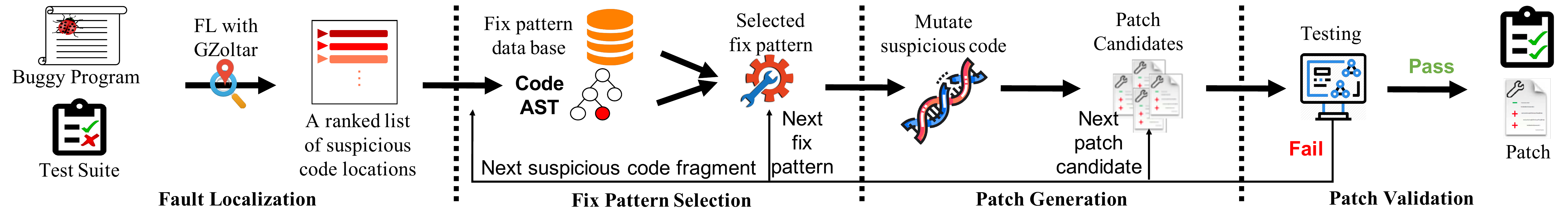}
    \caption{The overall workflow of TBar.}
    \label{fig:TBar}
\end{figure*}

%% file: tables/fpworks.tex
\begin{table}[!h]
	\centering
	\scriptsize
	\caption{Literature review on fix patterns for Java programs.}
	\resizebox{1\linewidth}{!}{
	\begin{threeparttable}
		\begin{tabular}{lcccc}
			\toprule
			\bf Authors & \bf APR tool name & \makecell[c]{\bf \# of fix\\\bf patterns} & \makecell[c]{\bf Publication\\\bf Venue} & \makecell[c]{\bf Publication\\\bf Year} \\
			\hline
			Pan et al.~\cite{pan2009toward} & - & 27 & EMSE & 2009\\
			Kim et al.~\cite{kim2013automatic} & PAR & 10 (16$^\ast$)& ICSE & 2013 \\
			Martinez et al.~\cite{martinez2016astor} & jMutRepair & 2 & ISSTA & 2016\\
			Durieux et al.~\cite{durieux2017dynamic} & NPEfix & 9 & SANER & 2017\\
			Long et al.~\cite{long2017automatic} & Genesis & 3 (108$^\ast$)& FSE & 2017 \\
			D. Le et al.~\cite{le2017s3} & S3 & 4 & FSE & 2017 \\
			Saha et al.~\cite{saha2017elixir} & ELIXIR & 8 (11$^\ast$) & ASE & 2017 \\
			Hua et al.~\cite{hua2018towards} & SketchFix & 6 & ICSE & 2018 \\
			Liu and Zhong~\cite{liu2018mining} & SOFix & 12 & SANER & 2018 \\
			Koyuncu et al.~\cite{koyuncu2018fixminer} & FixMiner & 28 & UL Tech Report & 2018\\
			Liu et al.~\cite{liu2018mining2} & - & 174 & TSE & 2018 \\
			Rolim et al.~\cite{rolim2018learning} & REVISAR & 9 & UFERSA Tech Report & 2018\\
			Liu et al.~\cite{liu2019avatar} & AVATAR & 13 & SANER & 2019\\
			\hline
			D. Le et al.~\cite{le2016history} & HDRepair$^\dagger$ & 11 & SANER & 2016\\
			Xin and Reiss~\cite{xin2017leveraging} & ssFix$^\dagger$ & 34 & ASE & 2017\\
			Wen et al.~\cite{wen2018context} & CapGen$^\dagger$ & 30 & ICSE & 2018\\
			Jiang et al.~\cite{jiang2018shaping} & SimFix$^\dagger$ & 16 & ISSTA & 2018\\
			\bottomrule
		\end{tabular}
		{$^\ast$In the PAR paper~\cite{kim2013automatic}, 10 fix patterns are presented, but 16 fix patterns are released online~\cite{parFP}. In Genesis, 108 code transformation schemas are inferred for three kinds of defects. In ELIXIR, there is one fix pattern that consists of four sub-fix patterns. 
		}
	\end{threeparttable}
	}
	\label{tab:FPPapers}
\end{table}

%% file: tables/fpQualitative.tex
\begin{table}[!tp]
	\centering
	\scriptsize
	\caption{Change properties of fix patterns.}
	\begin{threeparttable}
		\begin{tabular}{l|c|c|c|c}
			\toprule
			\makecell[l]{{\bf Fix}\\{\bf Pattern}} & \makecell[c]{{\bf Change}\\{\bf Action}} & \makecell[c]{{\bf Change}\\{\bf Graunlarity}} & {\bf Bug Context} & \makecell[c]{{\bf Change}\\{\bf Spread}} \\
			\hline
			FP1 & Insert & statement & cast expression & single\\\hline
			FP2.1 & \multirow{4}{*}{Insert} & \multirow{4}{*}{statement} & \multirow{4}{*}{\makecell[c]{a variable or\\ an expression \\returning non-\\primitive-type data}} & single\\\cline{1-1}\cline{5-5}
			\multirow{3}{*}{FP2.(2,3,4,5)} & &  & & \multirow{3}{*}{dual}\\
				 & &  & & \\
				  & &  & & \\\hline
			FP3 & Insert & statement & \makecell[c]{element access\\of array or\\collection variable} & single\\\hline
			FP4.(1,2,3,4) & Insert & statement & any statement & single\\\hline
			FP5 & Update & expression & \makecell[c]{class instance\\creation expression\\and clone method} & single\\\hline
			FP6.1 & Update & \multirow{3}{*}{expression} & \multirow{3}{*}{conditional expression} & \multirow{3}{*}{single}\\\cline{1-2}
			FP6.2 & Delete &  & & \\\cline{1-2}
			FP6.3 & Insert &  & &\\\hline
			FP7.1 & Update & expression & \makecell[c]{variable declaration\\expression} & single\\\hline
			FP7.2 & Update & expression & \makecell[c]{cast expression} & single\\\hline
			FP8.(1,2,3) & Update & expression & \makecell[c]{integral division\\expression} & single\\\hline
			FP9.(1,2) & Update & expression & literal expression & single\\\hline
			FP10.1 & \multirow{2}{*}{Update} & \multirow{4}{*}{\makecell[c]{expression,\\or statement}} & \multirow{4}{*}{\makecell[c]{method invocation,\\class instance creation,\\constructor, or\\super constructor}} & \multirow{4}{*}{single}\\\cline{1-1}
			FP10.2 &  &  & &\\\cline{1-2}
			FP10.3 & Delete &  & &\\\cline{1-2}
			FP10.4 & Insert &  & &\\\hline
			FP11.1 & Update & expression & \makecell[c]{assignment or\\infix-expression} & single\\\hline
			FP11.2 & Update & expression &  \makecell[c]{arithmetic\\infix-expression} & single\\\hline
			FP11.3 & Update & expression & instance of expression & single\\\hline
			FP12 & Update & expression & return statement & single	\\\hline
			FP13.(1, 2) & Update & expression & variable expression & single\\\hline
			FP14 & Move & statement & any statement &  \makecell[c]{single or\\multiple}\\\hline
			FP15.1 & Delete & statement & any statement &  \makecell[c]{single or\\multiple} \\\hline
			FP15.2 & Delete & method & any statement & multiple\\
			\bottomrule
		\end{tabular}
	\end{threeparttable}
	\label{tab:fpQualitative}
\end{table}

%% file: tables/fpActions.tex
\begin{table}[!tp]
	\centering
	\scriptsize
	\caption{Diversity of fix patterns w.r.t change properties.}
	\resizebox{1\linewidth}{!}{
	\begin{threeparttable}
		\begin{tabular}{lc|lc|lc}
			\toprule
			Action Type & \# fix patterns & Granularity & \# fix patterns & Spread & \# fix patterns \\
			\hline
			Update & 17 & Expression & 21 & Single- & \multirow{2}{*}{30}\\
			Delete & 4 & Statement & 17 & Statement & \\ \cline{5-6}
			Insert & 13 & Method & 1 & Multiple- & \multirow{2}{*}{7} \\
			Move & 1 &  & & Statements & \\
			\bottomrule
		\end{tabular}
	\end{threeparttable}
	}
	\label{tab:fpActions}
\end{table}

%% file: exp.tex
\section{Setup for Repair Experiments}
\label{sec:exp_setup}
In order to assess the effectiveness of fix patterns in the taxonomy presented in Section~\ref{sec:fp}, we design program repair experiments using the fix patterns as the main ingredients.
The produced APR system is then assessed on a widely-used benchmark in the repair community to allow reliable comparison against the state-of-the-art.

\subsection{\toolname: a Baseline APR System}
Based on the investigations of recurrently-used fix patterns, we build \toolname, a template-based APR tool which integrates the \nfixpatterns fix patterns presented in Section~\ref{sec:fp}.
We expect the APR community to consider \toolname as a baseline APR tool: new approaches must come up with novel techniques for solving auxiliary issues (e.g., repair precision, search space optimization, fault locations re-prioritization, etc.) to boost automated program repair beyond the performance that a straightforward application of common fix patterns can offer.
 Figure~\ref{fig:TBar} overviews the workflow that we have implemented in \toolname. We describe in the following subsections the role and operation of each process as well as all necessary implementation details.

\subsubsection{Fault Localization}
Fault localization is necessary for template-based APR as it allows to identify a list of suspicious code locations (i.e., buggy statements) on which to apply the fix patterns.
\toolname leverages the GZoltar~\cite{campos2012gzoltar} framework to automate the execution of test cases for each buggy program.
In this framework, we use the Ochiai~\cite{abreu2007accuracy} ranking metric to compute the suspiciousness scores of statements that are likely to be the faulty code locations.
This ranking metric has been demonstrated in several empirical studies~\cite{steimann2013threats,xie2013theoretical,xuan2014learning,pearson2017evaluating} to be effective for localizing faults in object-oriented programs.
The GZoltar framework for fault localization is also widely used in the literature of APR~\cite{martinez2016astor,xiong2017precise,xuan2017nopol,xin2017leveraging,wen2018context,koyuncu2018fixminer,kui2018live,jiang2018shaping,liu2019you,liu2019avatar}, allowing for a fair assessment of \toolname's performance against the state-of-the-art.

\subsubsection{Fix Pattern Selection}
In the execution of the repair pipeline, once the fault localization process yields a list of suspicious code locations,
\toolname sequentially attempts to select the encoded fix patterns from its database of fix patterns for each statement in the locations list.
The selection of fix patterns is conducted in a na\"ive way based on the AST context information of each suspicious statement.
Specifically, \toolname sequentially traverses each node of the suspicious statement AST from its first child node to its last leaf node and tries to match each node against the context AST of the fix pattern.
If a node can match any bug context presented in Table~\ref{tab:fpQualitative}, a related fix pattern will be matched to generate patch candidates with the corresponding code change pattern.
If the node is not a leaf node, \toolname keeps traversing its children nodes.
For example, if the first child node of a suspicious statement is a method invocation expression, it will be first matched with {\bf FP10. Mutate Method Invocation Expression} fix pattern. If the children nodes of the method invocation start from a variable reference, it will be matched with {\bf FP13. Mutate Variable} fix pattern as well. Other fix patterns follow the same manner.
After all expression nodes of a suspicious statement are matched with fix patterns,
\toolname further matches fix patterns from statement and method levels respectively.

\subsubsection{Patch Generation and Validation}
When a matching fix pattern is found (i.e., a pattern is selected for a suspicious statement), a patch is generated by mutating the statement, then the patched program is run against the test suite. If the patched program passes all tests successfully, the patch candidate is considered as a {\em plausible} patch~\cite{qi2015analysis}.
Once such a plausible patch is identified, \toolname stops generating other patch candidates for this bug to fix bugs in a standard and practical program repair workflow~\cite{martinez2016astor,xiong2017precise,xuan2017nopol,liu2019you,liu2019avatar}, but does not generate all plausible patches for each bug, unlike PraPR~\cite{ghanbari2018practical}.
Otherwise, the pattern selection and patch generation process is resumed until all AST nodes of buggy code are traversed. 
When several fix pattern contexts match one node, their actions are used for ordering: \toolname prioritizes Update over Insert that is over Delete, which is prioritized over Move.
In case of multiple donor code options for a given fix pattern, the candidate patches (each generated with a specific donor code) are ordered based on the distances between donor code node and buggy code node in the AST of the buggy code file:
priority is given to smaller distances.
Due to space limitation, detailed steps, illustrated in an algorithmic pseudo-code, are released in the replication package.

Considering that some buggy programs have several buggy locations, if a patch candidate can make a buggy program pass a sub-set of previously failing test cases without failing any previously passing test cases, this patch is considered as a plausible sub-patch of this buggy program. \toolname will further validate other patch candidates, 
until either a plausible patch is generated, or all patch candidates are validated, or \toolname exhausts the time limitation set (i.e., three hours) for repair attempts.

If a plausible patch is generated, we further manually check the equivalence between this patch and the ground-truth patch provided by developers and available in the Defects4J benchmark.
If the plausible patch is semantically equivalent to the ground-truth patch, the plausible patch is considered as {\em correct}. Otherwise, it is only considered as plausible.
We offer a replication package with extensive details on pattern implementation within \toolname. Source code is publicly available in the aforementioned GitHub repository.

\subsection{Assessment Benchmark}
For our empirical assessments, we selected the Defects4J~\cite{just2014defects4j} dataset as the evaluation benchmark of \toolname. This benchmark includes test cases for buggy Java programs with the associated developer fixes.
Defects4J is an ideal benchmark for the objective of this study, since it has been widely used by most recent state-of-the-art APR systems targeting Java program bugs.
Table~\ref{tab:defects} provides summary statistics on the bugs and test cases available in the version 1.2.0~\cite{defects4j}
of Defects4J which we use in this study.
\input{tables/benchmark.tex}

Overall, we note that, to date, {\bf 101} Defects4J bugs have been \underline{correctly} fixed by at least one APR tool published in the literature. Nevertheless, we recall that SimFix~\cite{jiang2018shaping} currently holds the record number of bugs fixed by a single tool, which is {\bf 34}.

%% file: tables/benchmark.tex

\begin{table}[!h]
	\centering
	\scriptsize
	\caption{Defects4J dataset information.}
	\resizebox{1\linewidth}{!}{
	\begin{threeparttable}
		\begin{tabular}{l|ccccccc}
			\toprule
			{\bf Project} & \makecell[c]{{\bf Chart}\\{\bf (C)}}  & \makecell[c]{{\bf Closure}\\{\bf(Cl)}} & \makecell[c]{{\bf Lang}\\{\bf(L)}} & \makecell[c]{{\bf Math}\\{\bf(M)}} & \makecell[c]{{\bf Mockito}\\{\bf(Mc)}} & \makecell[c]{{\bf Time}\\{\bf(T)}}& {\bf Total} \\
			\hline
			\# bugs & 26 & 133 & 65 & 106 & 38 & 27 & 395\\
			\# test cases &  2,205 &  7,927 &  2,245 &  3,602 & 1,457 &  4,130& 21,566 \\
			\hline
			\makecell[l]{\# fixed bugs by all\\APR tools (cf. \cite{liu2019you,liu2019avatar})}
			& 13 & 16 & 28 & 37 & 3 & 4 & 101\\
			\bottomrule
		\end{tabular}
	\end{threeparttable}
	}
	\label{tab:defects}
\end{table}

%% file: empiricalStudy.tex
\section{Assessment}
\label{sec:empiricalStudy}
This section presents and discusses the results of repair experiments with \toolname. In particular, we conduct two experiments for:
\begin{itemize}
	\item {\bf Experiment \#1:} Assessing the effectiveness of the various fix patterns implemented in \toolname. To avoid the bias that fault localization can introduce with its false positives (cf.~\cite{liu2019you}), we directly provide perfect localization information to \toolname.
	\item  {\bf Experiment \#2:} Evaluating \toolname in a normal program repair scenario. We investigate in particular the tendency of fix patterns to produce more or less incorrect patches. 
\end{itemize}

\subsection{Repair Suitability of Fix Patterns}
\label{sec:expI}

Our first experiment focuses on assessing the patch generation performance of fix patterns for real bugs. In particular, we investigate three research questions in Experiment \#1.
\questions{\footnotesize Research Questions for Experiment \#1}{\footnotesize
\begin{itemize}
\item[RQ1.] How many real bugs from Defects4J can be correctly fixed by fix patterns from our taxonomy?
\item[RQ2.] Can each Defects4J bug be fixed by different fix patterns?
\item[RQ3.] What are the properties of fix patterns that are successfully used to fix bugs?
\end{itemize}
}

In a recent study, Liu et al.~\cite{liu2019you} reported how fault localization techniques substantially affect
the repair performance of APR tools. Given that, in this experiment, the APR tool (namely \toolname) is only used as a means to apply the fix patterns in order to assess their effectiveness, we must eliminate the fault localization bias. Therefore, we assume that the bug positions at statement level are known, and we directly provide it to the patch generation step of \toolname, without running any fault localization tool (which is part of the normal APR workflow, see Figure~\ref{fig:TBar}).
To ensure readability across our experiments, we denote this version of the APR system  as $\toolname_p$ (where $p$ stands for \emph{\uline{p}erfect localization}). Table~\ref{tab:fixedBugs} summarizes the experimental results of $\toolname_p$.

\input{tables/fixedBugs.tex}

Among 395 bugs in the Defects4J benchmark, $\toolname_p$ can generate plausible patches for 101 bugs. 74 of these bugs are fixed with correct patches. We also note that $\toolname_p$ can
 partially fix\footnote{Partial fix: a patch makes the buggy program pass
a part of previously failed test cases without causing any new failed test cases~\cite{liu2019you}.}
20 bugs with plausible patches, and 8 of them are correct.
In a previous study, the {\tt kPAR}~\cite{liu2019you} baseline tool (i.e., a Java implementation of the PAR~\cite{kim2013automatic} seminal template-based APR tool) was correctly/plausibly fixing 36/55 Defects4J bugs when assuming perfect localization.

While the results of $\toolname_p$ are promising, $\sim$79\%(=314/395) of bugs cannot be correctly fixed with the available fix patterns. We manually investigated these unfixed bugs and make the following observations as research directions for improving the fix rates:
\begin{enumerate}[leftmargin=*]
	\item {\em Insufficient fix patterns.} Many bugs are not fixed by $\toolname_p$ simply due to the absence of matching fix patterns. This suggests that the fix patterns collected in the literature are far from being representative for real-world bugs. The community must thus keep contributing with effective techniques for mining fix patterns from existing patches.
	\item {\em Ineffective search of fix ingredients.} Template-based APR is
  a kind of search-based APR~\cite{wen2018context}: some fix patterns require donor code (i.e.,
  fix ingredients) to generate actual patches.
  For example, as shown in Figure~\ref{fig:chart10}, to apply the relevant fix pattern {\bf FP9.2}, one
  needs to identify fix ingredient ``{\tt ImageMapUtilities.htmlEs- cape}'' as the necessary in generating the patch.
  The current implementation of \toolname limits its search space for donor code to the ``{\bf\em local}'' file where the bug is localized. It is a limitation to find the correct donor code, but it reduces the risk of search space explosion. In addition, \toolname leverages the context of buggy code to prune away irrelevant fix ingredients. Therefore, some bugs
  cannot be fixed by \toolname although its fix pattern can match
  with code change actions. With more effective search strategies
  (e.g., larger search space such as fix ingredients from other projects as in ~\cite{kui2018live}),
  there might be more chances to fix more bugs.

 \end{enumerate}

\find{{\bf RQ1:} The collected fix patterns can be used to correctly fix 74 real bugs from the Defects4J dataset. A larger portion of the dataset remains however unfixed by $\toolname_p$, notably due to (1) the limitations of the fix patterns set and to (2) the na\"ive search strategy for finding relevant fix ingredients to build concrete patches from patterns.}

\begin{figure}[!ht]
    \centering
\begin{lstlisting}[language=diff,linewidth={\linewidth},frame=tb,basicstyle=\scriptsize\ttfamily]
   public String generateToolTipFragment(String toolTipText) {
-     return " title=\"" + toolTipText
+     return " title=\"" + ImageMapUtilities.htmlEscape(toolTipText) 
             + "\" alt=\"\"";
   }
   (@{\bf Code Change Action:}@)
   Replace variable "toolTipText" with a method invocation expression "ImageMapUtilities.htmlEscape(toolTipText)".
   (@{\bf Matchable fix pattern: FP9.2}@).
\end{lstlisting}
    \caption{Patch and code change action of fixing bug C-10.}
    \label{fig:chart10}
\end{figure}

\begin{figure}[!t]
	\centering
    \includegraphics[width=0.7\linewidth]{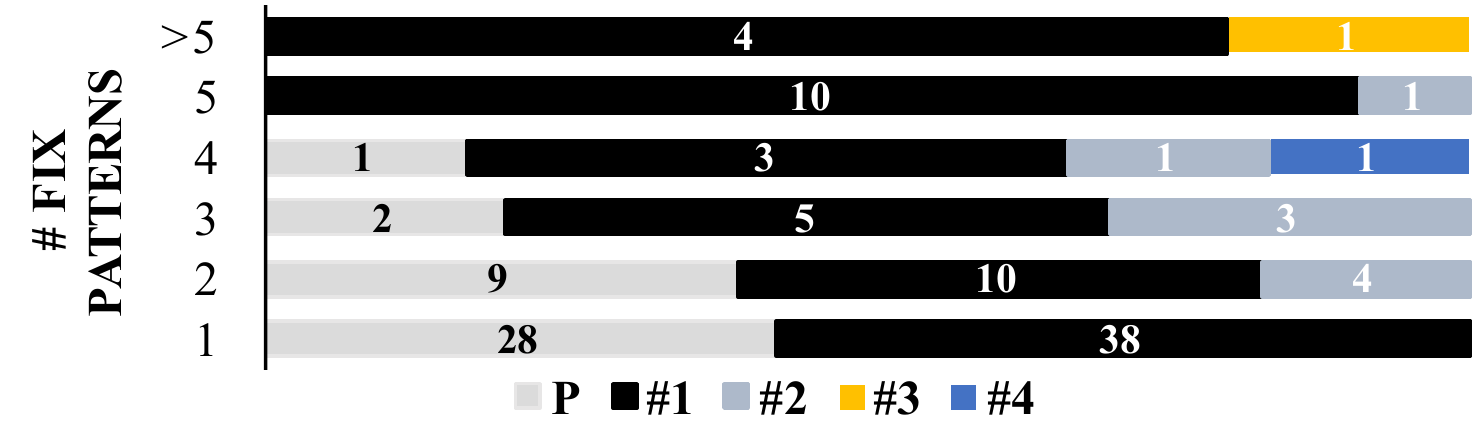}
    \caption{Number of bugs plausibly and correctly fixed by single or
    multiple fix patterns.}
    \label{fig:per:pattern}
\end{figure}

Figure~\ref{fig:per:pattern} summarizes the statistics on the number of
bugs that can be fixed by one or several fix patterns.
The {\em Y}-axis denotes
the number of fix patterns (i.e., $n=$ 1, 2, 3, 4, 5, and >5) that can generate
\emph{plausible} patches for a number of bugs ({\em X}-axis). The legend indicates that ``{\bf P}'' represents the number of plausible patches
generated by $\toolname_p$ (i.e.,  those that are not found to be correct).
``{\bf \#$k$}'', where $k\in[1,4]$, indicates that a bug can be correctly fixed by only $k$ fix patterns (although it may be plausibly fixed by more fix patterns).

Consider for the bottom-most bar in Figure~\ref{fig:per:pattern}: 66 (=28+38)
bugs can be plausibly fixed by a single pattern ({\em Y}-axis value is 1); it turns out that only 38 of them are correctly fixed.
Note that several patterns can generate (plausible) patches for a bug,
but not all patches are necessarily correct. For example, in the case of the top-most bar in Figure~\ref{fig:per:pattern}, 5 bugs are each plausibly fixed by over 5 fix patterns. However, only 1 bug is correctly fixed by 3 fix patterns.

In summary, 86\% (=$\frac{38 + 10 + 5 + 3 + 10 + 4}{74 + 7}$) of correctly fixed bugs (74 fully
and 7 partially fixed bugs) are exclusively fixed correctly by single patterns. In other words, generally, several fix patterns
can generate patches that can pass all test cases but, in most cases, the bug is correctly fixed by only one pattern. This finding suggests that
it is necessary to carefully select an appropriate fix pattern when attempting to fix a bug, in order to avoid plausible patches which may prevent the discovery of correct patches by halting the repair process (given that all tests are passing on the plausible patch).

\find{{\bf RQ2:} Some bugs can be plausibly fixed by different fix patterns. However, in most cases, only one fix pattern is adequate for generating a correct patch. This finding suggests a need for new research on fix pattern prioritization.}

Table~\ref{tab:exp1} details which bug is fixed by which fix pattern(s). We note that five fix patterns
(i.e., FP3, FP4.3, FP5, FP7.2 and FP11.3) cannot be used to generate a plausible patch for any Defects4J bug. Two fix patterns (i.e., FP9.2 and FP12) lead to plausible patches for some bugs, but none of them is correct. It does not necessarily suggest that the aforementioned fix patterns are useless (or ineffective) in APR. Instead, two reasons can explain their performance:
\begin{itemize}[leftmargin=*]
	\item The search for donor code may be inefficient for finding relevant ingredients for applying these patterns
	\item  The Defects4J dataset does not contain the types of bugs that can be addressed by these fix patterns.
\end{itemize}

In addition,
twenty (20) fix patterns lead to the generation of correct patches for some bugs. Most of these fix patterns are involved in the generation of plausible patches (which turn out to be incorrect). Interestingly, we found the cases of six (6) fix patterns which can generate several\footnote{Note that, in this experiment $\toolname_p$ generates and assesses all possible patch candidates for a given pair "bug location - fix pattern" with varying ingredients.} patch candidates, some which being correct and others being only plausible, for the same 10 bugs (as indicated in Table~\ref{tab:exp1} with `\LEFTcircle').
This observation further highlights the importance of selecting a relevant donor code for synthesizing patches: selecting an inappropriate donor code can lead to the generation of a plausible (but incorrect) patch, which will impede the generation of correct patches in a typical repair pipeline.

\find{Aside from fix patterns, fix ingredients collected in donor code are essential to be properly selected to avoid patches that are plausible but may yet be incorrect.}

%
%

\begin{figure}[!h]
	\centering
    \includegraphics[width=\linewidth]{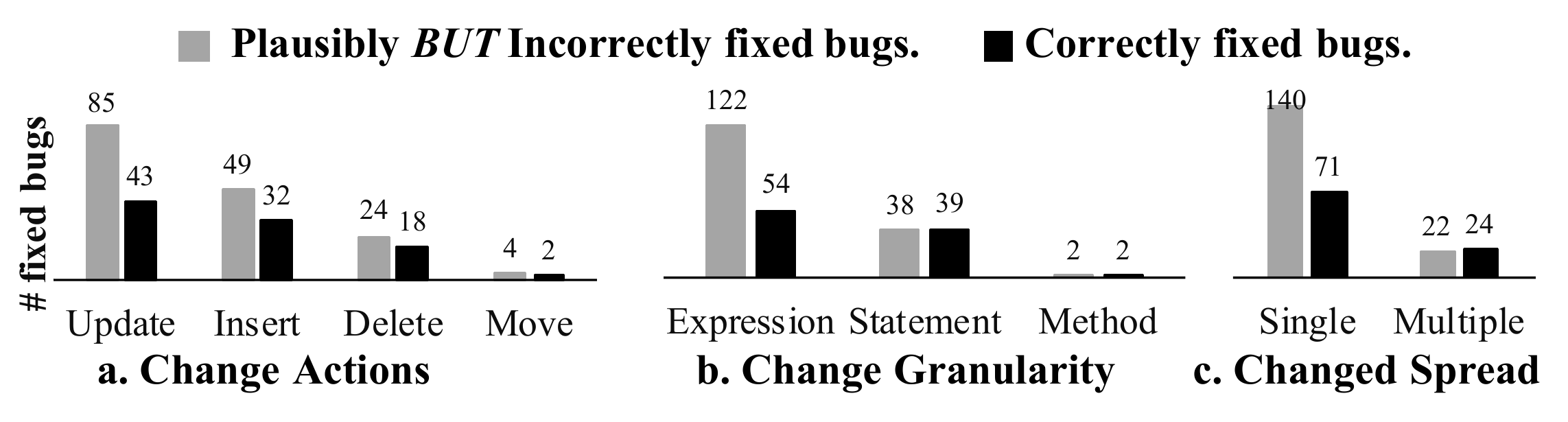}
    \caption{Qualitative statistics of bugs fixed by fix patterns.}
    \label{fig:sFB}
\end{figure}

We further inspect properties of fix patterns, such as change actions,
granularity, and the number of changed statements in patches.
The statistics are shown in Figure~\ref{fig:sFB}, highlighting the number of plausible (but incorrect) and correct patches for the different property dimensions through which fix patterns can be categorized.


More bugs are fixed by {\em Update} change actions than any by any other actions.
Similarly, fix patterns targeting expressions fix more bugs correctly than patterns targeting statements and methods. However, fix patterns mutating whole statements have a higher rate of correct patches among their plausible generated patches.
Finally, fix patterns changing only single statements
can correctly fix more bugs than those touching multiple statements. Fix patterns targeting multi-statements have however a higher rate of correctness.

\find{{\bf RQ3:} There are noticeable differences between successful repair among fix patterns depending on their properties related to implemented change actions, change granularity and change spread.
}

\input{tables/exp_1_Results.tex}

\input{tables/exp2FixedBugs.tex}

\input{tables/fpFixedBugs.tex}

\subsection{Repair Performance Comparison: \toolname vs State-of-the-art APR tools}
\label{sec:expII}

Our second experiment evaluates \toolname in a realistic setting for patch generation, allowing for reliable comparison against the state-of-the-art in the literature. Concretely, we investigate two research questions in Experiment \#2.

\questions{\footnotesize Research Questions for Experiment \#2}{\footnotesize
\begin{itemize}
\item[RQ4.] What performance can be achieved by \toolname in a standard and practical repair scenario?
\item[RQ5.] To what extent are the different fix patterns sensitive to noise in fault localization (i.e., spotting buggy code locations)?
\end{itemize}
}

In this experiment we implement a realistic scenario, using a normal fault localization (i.e., no assumption of perfect localization as for $\toolname_p$) on Defects4J bugs. To enable a fair comparison with performance results recorded in the literature, \toolname leverages
a standard configuration in the literature~\cite{liu2019you} with GZoltar~\cite{campos2012gzoltar}
and Ochiai~\cite{abreu2007accuracy}.
Furthermore, \toolname does not utilize any additional technique to improve
the accuracy of fault localization,
such as crashed stack trace (used by ssFix~\cite{xin2017leveraging}), predicate
switching~\cite{zhang2006locating} (used by ACS~\cite{xiong2017precise}), or test case
purification~\cite{xuan2014test} (used by SimFix~\cite{jiang2018shaping}).


With respect to the patch generation step, contrary to the experiment with $\toolname_p$ where all positions of multi-locations bugs were known (cf. Section~\ref{sec:expI}), \toolname adapts a ``first-generated and first-selected'' strategy to progressively apply fix patterns, one at a time, in various suspicious code locations:
\toolname generates a patch $p_i$, using a fix pattern that matches a given bug.
If $p_i$ passes a subset of
previously-failing test cases without failing any previously-passing test case, \toolname selects $p_i$ as a plausible patch for the bug.
Then, \toolname continues to validate another patch $p_{i+1}$ (which can be generated by the same fix pattern on the same code entity with other ingredients, or on another code location).
When $p_{i+1}$ passes a subset of test cases as $p_i$, if $p_{i+1}$ is generated for the same buggy code entity as $p_i$, $p_{i+1}$ will be abandoned;
otherwise, \toolname takes $p_{i+1}$ as another plausible patch as well.
Through this process, \toolname creates a patch set $P$ = \{ $p_i$, $p_{i+1}$, ...\} of plausible patches.
Here, as soon as any patch can pass all the given test cases for a given bug,
\toolname takes it as a plausible patch for the given bug, which is
regarded as a \emph{fully-fixed} bug, and all $p_i \in P$ will be abandoned. Otherwise,
our tool yields $P$, a set of plausible patches that can each partially fix the given bug.


We run the \toolname APR system against the buggy programs of the Defects4J dataset. Table~\ref{tab:comparison} presents the performance of \toolname in comparison with recent state-of-the-art APR tools from the literature. \toolname can fix 81 bugs with plausible patches, 43 of which are correctly fixed. No other APR tool had reached this number of fixed bugs. Nevertheless, its precision (ratio of correct vs. plausible patches) is lower than some recent tools such as CapGen and SimFix which employs sophisticated techniques to select fix ingredients.
Nonetheless, it is noteworthy that, despite using fix patterns catalogued in the literature, we can fix three bugs (namely Cl-86,L-47,M-11) which had never been fixed by any APR system: M-11 is fixed by a pattern found by a standalone fix pattern mining tool~\cite{liu2018mining2} but which was not encoded by any APR system yet. Cl-86 and L-47 are fixed by patterns that were not applied to Defects4J.

\find{{\bf RQ4:} \toolname outperforms all recent state-of-the-art APR tools that were evaluated on the Defects4J dataset. It correctly fixes 43 bugs, while the runner-up (SimFix) is reported to correctly fix 34 bugs.}

It is noteworthy that \toolname performs significantly less than $\toolname_p$ (43 vs. 74 correctly fixed bugs). This result is in line with a recent study~\cite{liu2019you}, which demonstrated that fault localization imprecision is detrimental to APR repair performance.
Table~\ref{tab:exp1} summarizes information about the number of bugs each fix pattern contributed to fixing with $\toolname_p$. While only 4 fix patterns did not lead to the generation of any plausible patch when assuming perfect localization. With \toolname, it is the case for 13 fix patterns (see Table~\ref{tab:fpFixedBugs}). This observation further confirms the impact of fault localization noise.



\begin{figure}[!h]
	\centering
    \includegraphics[width=0.8\linewidth]{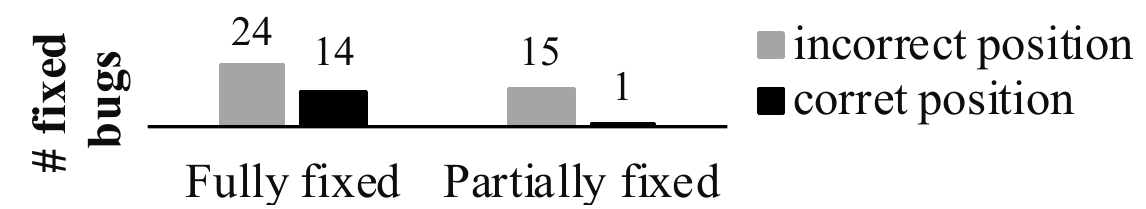}
    \caption{The mutated code positions of plausibly but incorrectly fixed bugs.}
    \label{fig:pVSp}
\end{figure}

We propose to examine the locations where \toolname applied fix patterns to generate plausible but incorrect patches. As shown in Figure~\ref{fig:pVSp}, \toolname has made changes on incorrect positions (i.e., non-buggy locations) for 24 out of the 38 fully-fixed and 15 out of the 16 partially-fixed bugs.


Even when \toolname applies a fix pattern to the precise buggy location, the generated patch may be incorrect. As shown in Figure~\ref{fig:pVSp}, 14 patches that fully fix Defects4J bugs
mutate the correct locations: in 3 cases, the fix patterns were inappropriate; in 2 other cases, \toolname failed to locate relevant donor code; for the remaining, \toolname does not support the required fix patterns.

\begin{figure}[!h]
	\centering
	\includegraphics[width=\linewidth]{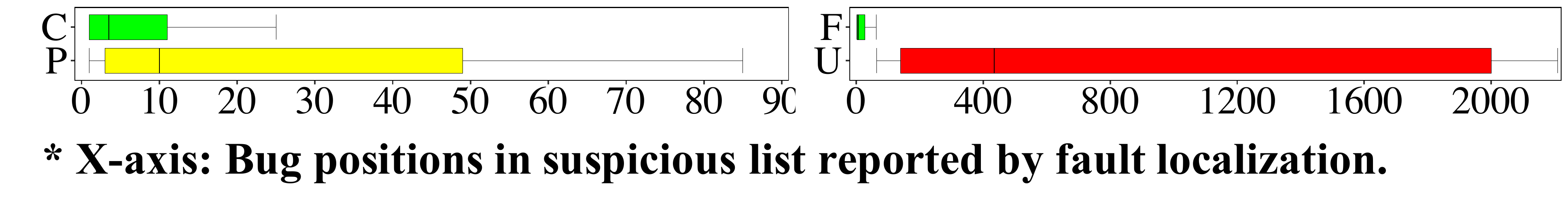}
	\caption{Distribution of the positions of buggy code locations in fault localization list of suspicious statements. {\small {\em C} and {\em P} denote Correctly- and Plausibly- (but incorrectly) fixed bugs, respectively. {\em F} and {\em U} denote Fixed and Unfixed bugs.} }
    \label{fig:distPositions}
\end{figure}

Finally, Figure~\ref{fig:distPositions} illustrates the impact of fault localization performance: unfixed bugs (but correctly fixed by $\toolname_p$) are generally more poorly localized than correctly fixed bugs. Similarly, we note that many plausible but incorrect patches are generated for bugs which are not well localized (i.e., several false positive buggy locations are mutated leading to plausible but incorrect patches).

Average positions bugs (in fault localization suspicious list) are also provided in Table~\ref{tab:fpFixedBugs}. It appears that some fix patterns (e.g., FP2.1,  FP6.3, FP10.2) can correctly fix bugs that are poorly localized, showing less sensitivity to fault localization noise than others.

\find{{\bf RQ5:} Fault localization noise has a significant impact on the performance of \toolname. Fix patterns are diversely sensitive to the false positive locations that are recommended as buggy positions.}






%

%% file: tables/fixedBugs.tex
\begin{table}[!ht]
	\centering
	\scriptsize
	\caption{Number of bugs fixed by fix patterns with $\toolname_p$.}
	\label{tab:fixedBugs}
	\resizebox{1.0\linewidth}{!} {
	\begin{threeparttable}
		\begin{tabular}{l|cccccc|c}
			\toprule
			{\bf Fixed Bugs} & {\bf  C} &  {\bf  Cl} & {\bf L} & {\bf M} & {\bf Mc} & {\bf T} & {\bf Total} \\ \hline
			\# of Fully Fixed Bugs & 12/13   & 20/26 & 13/18  & 23/35 & 3/3 & 3/6 & 74/101\\
			\# of Partially Fixed Bugs & 2/4  & 3/6  & 1/4  & 0/4   & 0/0  & 1/1 & 7/20\\
			\bottomrule
		\end{tabular}
		{\scriptsize$^\ast$We provide $x/y$ numbers: $x$ is the number of correctly fixed bugs; $y$ is the number of bugs fixed with plausible patches. The same notation applies to Table~\ref{tab:comparison}.}
	\end{threeparttable}
	}
\end{table}

%% file: tables/exp_1_Results.tex
\begin{table*}[!h]
	\centering
	\scriptsize
	\caption{Defects4j bugs fixed by fix patterns.}
	\label{tab:exp1}
	\resizebox{1.0\linewidth}{!} {\scriptsize
	\begin{threeparttable}
		\begin{tabular}{l|c|c|c|c|c|c|c|c|c|c|c|c|c|c|c|c|c|c|c|c|c|c|c|c|c|c|c|c|c|c|c|c|c|c|c||c}
			\toprule
			\multirow{2}{*}{\makecell[l]{{\bf Bug}\\{\bf ID}}} & \multirow{2}{*}{\rotatebox[origin=l]{90}{\bf FP1}} & \multicolumn{5}{c|}{\bf FP2} & \multirow{2}{*}{\rotatebox[origin=l]{90}{\bf FP3}} & \multicolumn{4}{c|}{\bf FP4} & \multirow{2}{*}{\rotatebox[origin=l]{90}{\bf FP5}} & \multicolumn{3}{c|}{\bf FP6} & \multicolumn{2}{c|}{\bf FP7} & \multicolumn{3}{c|}{\bf FP8} & \multicolumn{2}{c|}{\bf FP9} & \multicolumn{4}{c|}{\bf FP10}& \multicolumn{3}{c|}{\bf FP11} & \multirow{2}{*}{\rotatebox[origin=l]{90}{\bf FP12}} & \multicolumn{2}{c|}{\bf FP13} & \multirow{2}{*}{\rotatebox[origin=l]{90}{\bf FP14}} & \multicolumn{2}{c||}{\bf FP15} & \\\cline{3-7}\cline{9-12}\cline{14-30}\cline{32-33}\cline{35-36}
			 & & 1 & 2 & 3 & 4 & 5 & & 1 & 2 & 3 & 4 & & 1 & 2 & 3 & 1 & 2 & 1 & 2 & 3 & 1 & 2 & 1 & 2 & 3 & 4 & 1 & 2 & 3 & & 1 & 2 & & 1 & 2 & \\			\hline
\rowcolor{lightgray}C-1	 &  &  &  &  &  &  &  &  &  &  &  &  &  &  & \ding{109}	 &  &  &  &  &  &  &  &  &  &  &  & \ding{108}	 &  &  &  &	 & \ding{109} & \ding{109} & \ding{109}	 &  & 1/5 \\
C-4	 &  & \ding{108}	 & \ding{109}	 &  & \ding{108}	 &  &  &  &  &  &  &  &  &  &  &  &  &  &  &  &  &  &  &  &  &  &  &  &  &  &  &  &  &  &  & 2/3 \\
\rowcolor{lightgray}C-7	 &  &  &  &  &  &  &  &  &  &  & \ding{109}	 &  &  &  &  &  &  &  &  &  &  &  &  &  &  &  &  &  &  &  & \LEFTcircle	 &  &  &  &  & 1/2 \\
C-8	 &  &  &  &  &  &  &  &  &  &  &  &  &  &  &  &  &  &  &  &  &  &  &  &  &  &  &  &  &  &  & \ding{108}	 &  &  &  &  & 1/1 \\
\rowcolor{lightgray}C-9	 &  &  &  &  &  &  &  &  &  &  &  &  &  &  & \ding{108}	 &  &  &  &  &  &  & \ding{109}	 &  &  &  &  &  &  &  &  &  &  &  &  &  & 1/2 \\
C-11	 &  &  &  &  &  &  &  &  &  &  &  &  &  &  &  &  &  &  &  &  &  &  &  &  &  &  &  &  &  &  & \ding{108}	 &  &  &  &  & 1/1 \\
\rowcolor{lightgray}C-12	 &  &  &  &  &  &  &  & \ding{108}	 &  &  &  &  &  &  &  &  &  &  &  &  &  &  &  &  &  &  &  &  &  &  &  &  &  &  &  & 1/1 \\
C-14	 &  &  & \ding{108}	 & \ding{108}	 &  &  &  &  &  &  &  &  &  &  &  &  &  &  &  &  &  &  &  &  &  &  &  &  &  &  & \uline{\ding{109}}	 &  &  &  &  & 2/3 \\
\rowcolor{lightgray}C-18	 &  &  &  &  &  &  &  & \uline{\ding{109}}	 &  &  &  &  &  &  & \uline{\ding{109}}	 &  &  &  &  &  &  &  &  &  &  &  & \uline{\ding{109}}	 &  &  &  & \uline{\ding{109}}	 &  &  & \uline{\ding{108}}	 &  & 1/5 \\
C-19	 &  &  &  &  &  & \ding{108}	 &  &  &  &  &  &  &  &  &  &  &  &  &  &  &  &  &  &  &  &  &  &  &  &  &  &  &  &  &  & 1/1 \\
\rowcolor{lightgray}C-20	 &  &  &  &  &  &  &  &  &  &  &  &  &  &  &  &  &  &  &  &  &  &  &  & \ding{108}	 &  &  &  &  &  &  &  &  &  &  &  & 1/1 \\
C-24	 &  &  &  &  &  &  &  &  &  &  &  &  &  &  &  &  &  &  &  &  &  &  &  &  &  &  &  &  &  &  & \ding{108}	 &  &  &  &  & 1/1 \\
\rowcolor{lightgray}C-25	 &  &  & \uline{\ding{108}}	 &  &  &  &  &  &  &  & \uline{\ding{109}}	 &  &  &  &  &  &  &  &  &  &  &  &  &  &  &  &  &  &  &  & \uline{\ding{109}}	 &  &  &  &  & 1/3 \\
C-26	 &  &\ding{108}  & \ding{109}	 & \ding{108}	 & 	 &  &  &  &  &  &  &  &  &  &  &  &  &  &  &  &  &  &  &  &  &  &  &  &  &  &  &  &  &  &  & 2/3 \\
\hline
\rowcolor{lightgray}Cl-2	 &  & \ding{108} & \ding{109}	 & \ding{109}	 &	 &  &  &  &  &  &  &  &  &  &  &  &  &  &  &  &  &  &  &  &  &  &  &  &  &  &  &  &  &  &  & 1/3 \\
Cl-4	 &  & & & 	 &	 &  &  &  &  &  &  &  &  &  &  &  &  &  &  &  &  &  &  \ding{108}&  &   &  &  &  &  &  &  &  &  &  &  & 1/1 \\
\rowcolor{lightgray}Cl-6	 &  &  &  &  &  &  &  &  &  &  &  &  &  & \uline{\ding{109}}	 & \uline{\ding{109}}	 &  &  &  &  &  &  &  &  &  &  &  & \uline{\ding{109}}	 &  &  &  & \uline{\ding{109}}	 & \uline{\ding{109}}	 &  & \uline{\ding{108}}	 &  & 1/6 \\
Cl-10	 &  &  &  &  &  &  &  &  &  &  &  &  &  &  &  &  &  &  &  &  &  &  & \ding{108}	 &  &  &  &  &  &  &  &  &  &  &  &  & 1/1 \\
\rowcolor{lightgray}Cl-11	 &  &  &  &  &  &  &  &  &  &  &  &  & \ding{109}	 &  & \ding{109}	 &  &  &  &  &  &  &  &  &  &  &  & \ding{109}	 &  &  &  & \ding{109}	 &  &  & \ding{108}	 &  & 1/5 \\
Cl-13	 &  &  &  &  &  &  &  &  &  &  &  &  &  &  &  &  &  &  &  &  &  &  &  &  &  &  &  &  &  &  &  &  & \ding{108}	 &  &  & 1/1 \\
\rowcolor{lightgray}Cl-18	 &  &  &  &  &  &  &  &  &  &  &  &  &  & \ding{108}	 &  &  &  &  &  &  &  &  &  &  &  &  & \ding{109}	 &  &  &  &  &  &  &  &  & 1/2 \\
Cl-21	 &  &  &  &  &  &  &  &  &  &  &  &  & \ding{109}	 &  & \ding{109}	 &  &  &  &  &  &  &  &  &  &  &  & \ding{109}	 &  &  &  & \ding{109}	 &  &  & \ding{108}	 &  & 1/5 \\
\rowcolor{lightgray}Cl-22	 &  &  &  &  &  &  &  &  &  &  &  &  & \ding{109}	 &  & \ding{109}	 &  &  &  &  &  &  &  &  &  &  &  & \ding{109}	 &  &  &  & \ding{109}	 &  &  & \ding{108}	 &  & 1/5 \\
Cl-31	 &  &  &  &  &  &  &  &  &  &  &  &  &  & \ding{108}	 &  &  &  &  &  &  &  &  &  &  &  &  & \ding{109}	 &  &  &  &  &  &  &  &  & 1/2 \\
\rowcolor{lightgray}Cl-38	 &  &  &  &  &  &  &  &  &  &  &  &  & \ding{109}	 & \ding{109}	 &  &  &  &  &  &  &  &  &  &  &  &  & \ding{108}	 &  &  &  &  &  &  &  &  & 1/3 \\
Cl-40	 &  &  &  &  &  &  &  &  &  &  &  &  &  &  &  &  &  &  &  &  & \ding{108}	 &  &  &  &  &  &  &  &  &  &  &  &  &  &  & 1/1 \\
\rowcolor{lightgray}Cl-46	 &  &  &  &  &  &  &  &  &  &  &  &  &  &  &  &  &  &  &  &  &  &  &  &  &  &  &  &  &  &  &  &  &  &  & \ding{108}	 & 1/1 \\
Cl-62	 &  &  &  &  &  &  &  &  &  &  &  &  & \ding{109}	 & \ding{109}	 & \ding{109}	 &  &  &  &  &  &  &  &  &  &  &  & \LEFTcircle	 &  &  &  &  &  &  & \ding{109}	 &  & 1/5 \\
\rowcolor{lightgray}Cl-63	 &  &  &  &  &  &  &  &  &  &  &  &  & \ding{109}	 & \ding{109}	 & \ding{109}	 &  &  &  &  &  &  &  &  &  &  &  & \LEFTcircle	 &  &  &  &  &  &  & \ding{109}	 &  & 1/5 \\
Cl-70	 &  &  &  &  &  &  &  &  &  &  &  &  &  &  &  &  &  &  &  &  & \ding{108}	 &  &  &  &  &  &  &  &  &  &  &  &  &  &  & 1/1 \\
\rowcolor{lightgray}Cl-73	 &  &  &  &  &  &  &  &  &  &  &  &  &  &  &  &  &  &  &  &  &  &  &  &  &  &  & \ding{108}	 &  &  &  &  &  &  &  &  & 1/1 \\
Cl-85	 &  &  &  &  &  &  &  &  &  &  &  &  &  &  &  &  &  &  &  &  &  &  &  &  &  &  &  &  &  &  &  &  &  & \uline{\ding{108}}	 &  & 1/1 \\
\rowcolor{lightgray}Cl-86	 &  &  &  &  &  &  &  &  &  &  &  &  &  &  &  &  &  &  &  &  & \ding{108}	 &  &  &  &  &  &  &  &  &  &  &  &  &  &  & 1/1 \\
Cl-102	 &  &  &  &  &  &  &  & \ding{108}	 &  &  &  &  &  &  &  &  &  &  &  &  &  &  &  &  &  &  &  &  &  &  &  &  & \ding{108}	 &  &  & 2/2 \\
\rowcolor{lightgray}Cl-106	 &  &  &  &  &  &  &  &  &  &  &  &  &  &  &  &  &  &  &  &  &  &  &  &  &  &  &  &  &  &  &  & \uline{\ding{109}}	 &  & \uline{\ding{108}}	 &  & 1/2 \\
Cl-115	 &  &  &  &  &  &  &  &  &  &  &  &  & \ding{109}	 &  & \ding{109}	 &  &  &  &  &  &  &  &  &  &  &  &  &  &  &  & \ding{109}	 &  & \uline{\ding{109}}	 & \ding{108}	 &  & 1/5 \\
\rowcolor{lightgray}Cl-126	 &  &  & \uline{\ding{109}}	 & \uline{\ding{109}} & 	 &  &  &  &  &  &  &  & \ding{109}	 &  & \ding{109}	 &  &  &  &  &  &  &  &  &  &  &  &  &  &  &  & \ding{109}	 &  &  & \ding{108}	 &  & 1/6 \\
\hline
L-6	 &  &  &  &  &  &  &  &  &  &  &  &  &  &  &  &  &  &  &  &  &  &  &  &  &  &  &  &  &  &  & \ding{108}	 &  &  &  &  & 1/1 \\
\rowcolor{lightgray}L-7	 &  &  &  &  &  &  &  &  &  &  &  &  & \ding{109}	 &  & \ding{109}	 &  &  &  &  &  &  &  &  &  &  &  &  &  &  &  & \ding{109}	 &  &  & \ding{108}	 &  & 1/4 \\
L-10	 &  &  &  &  &  &  &  &  &  &  &  &  &  &  & \LEFTcircle	 &  &  &  &  &  &  &  &  &  &  &  &  &  &  &  &  &  &  & \ding{108}	 &  & 2/2 \\
\rowcolor{lightgray}L-15	 &  &  &  &  &  &  &  &  &  &  &  &  & \uline{\ding{109}}	 & \uline{\ding{108}}	 & \uline{\ding{109}}	 &  &  &  &  &  &  &  &  &  &  &  & \uline{\ding{109}}	 &  &  &  &  &  &  & \uline{\ding{109}}	 &  & 1/5 \\
L-22	 &  &  &  &  &  &  &  &  &  &  &  &  &  & \ding{109}	 & \ding{109}	 &  &  &  &  &  &  &  &  &  &  &  & \LEFTcircle	 &  &  &  & \ding{109}	 &  &  & \ding{109}	 &  & 1/5 \\
\rowcolor{lightgray}L-24 &  &  &  &  &  &  &  &  &  &  &  &  &  & & \ding{108} &  &  &  &  &  &  &  &  &  &  &  &  &  &  &  &  &  &  &  &  & 1/1 \\
L-26 &  &  &  &  &  &  &  &  &  &  &  &  &  & &  &  &  &  &  &  &  &  &  &  &  & \ding{108} &  &  &  &  &  &  &  &  &  & 1/1 \\
\rowcolor{lightgray}L-33	 &  & \ding{108} &  &  & 	 &  &  &  &  &  &  &  &  &  &  &  &  &  &  &  &  &  &  &  &  &  &  &  &  &  &  &  &  &  &  & 1/1 \\
L-39	 &  & \ding{108} &  &  & 	 &  &  &  &  &  & \ding{109}	 &  &  &  &  &  &  &  &  &  &  &  &  &  &  &  &  &  &  &  &  & \ding{109}	 &  &  &  & 1/3 \\
\rowcolor{lightgray}L-47	 &  &  &  & \ding{108}	 &  &  &  &  &  &  &  &  &  &  &  &  &  &  &  &  &  &  &  &  &  &  &  &  &  &  &  &  &  &  &  & 1/1 \\
L-51	 &  &  &  &  &  &  &  &  & \ding{108}	 &  &  &  &  &  &  &  &  &  &  &  &  &  &  &  &  &  &  &  &  &  &  &  &  &  &  & 1/1 \\
\rowcolor{lightgray}L-57	 &  & \ding{109} & \ding{109}	 &  & 	 &  &  &  &  &  &  &  &  &  &  &  &  &  &  &  &  &  &  &  &  &  &  &  &  & \uline{\ding{109}}	 & \ding{108}	 & \ding{108}	 &  &  &  & 2/5 \\
L-59	 &  &  &  &  &  &  &  &  &  &  &  &  &  &  &  &  &  &  &  &  &  &  &  &  &  &  &  &  &  &  & \ding{108}	 &  &  &  &  & 1/1 \\
\rowcolor{lightgray}L-63	 &  &  &  &  &  &  &  &  &  &  &  &  &  &  & \ding{109}	 &  &  &  &  &  &  &  &  & \ding{109}	 &  &  & \ding{109}	 &  &  &  & \ding{109}	 & \ding{109}	 & \ding{109}	 & \ding{108}	 &  & 1/7 \\
\hline
M-4	 &  &  & \ding{108}	 &  &  &  &  &  &  &  &  &  &  &  &  &  &  &  &  &  &  &  &  &  &  &  &  &  &  &  &  &  &  &  &  & 1/1 \\
\rowcolor{lightgray}M-5	 &  &  &  &  &  &  &  &  &  &  &  &  &  &  &  &  &  &  &  &  &  &  &  &  &  &  &  &  &  &  & \ding{108}	 & \ding{109}	 &  &  &  & 1/2 \\
M-11	 &  &  &  &  &  &  &  &  &  &  &  &  &  &  &  &  &  & \ding{108}	 & \ding{108}	 & \ding{108}	 & \ding{108}	 &  &  &  &  &  &  &  &  &  &  &  &  &  &  & 4/4 \\
\rowcolor{lightgray}M-15	 &  &  &  &  &  &  &  &  &  &  &  &  &  &  & \LEFTcircle	 &  &  &  &  &  &  &  &  &  &  &  &  &  &  &  &  &  &  &  &  & 1/1 \\
M-22	 &  &  &  &  &  &  &  &  &  &  &  &  &  &  &  &  &  &  &  &  & \ding{108}	 & \uline{\ding{109}}	 &  &  &  &  &  &  &  &  &  &  &  &  &  & 1/2 \\
\rowcolor{lightgray}M-30	 &  &  &  &  &  &  &  &  &  &  &  &  &  &  &  & \LEFTcircle	 &  &  &  &  &  &  &  &  &  &  &  &  &  &  &  &  &  &  &  & 1/1 \\
M-33	 &  &  &  &  &  &  &  &  &  &  &  &  &  &  & \ding{109}	 &  &  &  &  &  &  & \ding{109}	 &  &  &  &  &  &  &  &  & \ding{108}	 &  &  &  &  & 1/3 \\
\rowcolor{lightgray}M-34	 &  &  &  &  &  &  &  &  &  &  &  &  &  &  &  &  &  &  &  &  &  &  &  &  &  &  &  &  &  &  &  & \ding{108}	 &  &  &  & 1/1 \\
M-35	 &  &  &  &  &  &  &  & \ding{108}	 &  &  &  &  &  &  &  &  &  &  &  &  &  &  &  &  &  &  &  &  &  &  &  &  &  &  &  & 1/1 \\
\rowcolor{lightgray}M-50	 &  &  &  &  &  &  &  &  &  &  & \ding{109}	 &  & \ding{109}	 &  & \ding{109}	 &  &  &  &  &  &  & \ding{109}	 &  &  &  &  & \ding{109}	 & \ding{109}	 &  &  & \ding{109}	 & \ding{109}	 &  & \ding{108}	 &  & 1/9 \\
M-57	 &  &  &  &  &  &  &  &  &  &  &  &  &  &  &  & \ding{108}	 &  &  &  &  &  &  &  &  &  &  &  &  &  &  &  &  &  &  &  & 1/1 \\
\rowcolor{lightgray}M-58	 &  &  &  &  &  &  &  &  &  &  &  &  &  &  &  & 	 &  &  &  &  &  &  &  &  &  \ding{108}&  &  &  &  &  &  &  &  &  &  & 1/1 \\
M-59	 &  &  &  &  &  &  &  &  &  &  &  &  &  &  &  &  &  &  &  &  &  &  &  &  &  &  &  &  &  &  & \ding{108}	 & \ding{109}	 &  &  &  & 1/2 \\
\rowcolor{lightgray}M-65	 &  &  &  &  &  &  &  &  &  &  &  &  &  &  &  &  &  &  &  &  &  &  &  &  &  &  & \ding{108} &  &  &  &  &  & 	 &  &  & 1/1 \\
M-70  &  &  &  &  &  &  &  &  &  &  &  &  &  &  &  & 	 &  &  &  &  &  &  &  &  &  & \ding{108} &  &  &  &  &  &  &  &  &  & 1/1 \\
\rowcolor{lightgray}M-75	 &  &  &  &  &  &  &  &  &  &  &  &  &  &  &  &  &  &  &  &  &  &  & \ding{108}	 &  &  &  &  &  &  &  &  &  &  &  &  & 1/1 \\
M-77	 &  &  &  &  &  &  &  &  &  &  & \uline{\ding{109}}	 &  &  &  &  &  &  &  &  &  &  &  &  &  &  &  & \uline{\ding{108}}	 &  &  &  & \uline{\ding{109}}	 &  &  &  & \uline{\ding{108}}	 & 2/4 \\
\rowcolor{lightgray}M-79	 &  &  &  &  &  &  &  &  &  &  &  &  &  &  &  & \LEFTcircle	 &  &  &  &  &  &  &  &  &  &  &  &  &  &  &  &  &  &  &  & 1/1 \\
M-80	 &  &  &  &  &  &  &  &  &  &  &  &  &  &  &  &  &  &  &  &  &  & \ding{109}	 &  &  &  &  &  & \ding{108}	 &  &  & \ding{109}	 & \ding{109}	 &  &  &  & 1/4 \\
\rowcolor{lightgray}M-82	 &  &  &  &  &  &  &  &  &  &  & \ding{109}	 &  &  &  & \ding{109}	 &  &  &  &  &  &  & \ding{109}	 &  &  &  &  & \ding{108}	 &  &  &  & \ding{109}	 &  &  &  &  & 1/5 \\
M-85	 &  &  &  &  &  &  &  &  &  &  & \ding{109}	 &  & \LEFTcircle	 &  & \LEFTcircle	 &  &  &  &  &  &  & \ding{109}	 &  &  &  &  & \ding{108}	 &  &  &  & \ding{109}	 & \ding{109}	 &  & \ding{109}	 &  & 3/8 \\
\rowcolor{lightgray}M-89	 & \ding{108}	 &  &  &  &  &  &  &  &  &  &  &  &  &  &  &  &  &  &  &  &  &  &  &  &  &  &  &  &  &  &  &  &  &  &  & 1/1 \\
M-98	 &  &  &  &  &  &  &  &  &  &  &  &  &  &  &  &  &  &  &  &  &  &  &  &  &  &  &  &  &  &  & \ding{108}	 &  &  &  &  & 1/1 \\
\hline
\rowcolor{lightgray}Mc-26	 &  &  &  &  &  &  &  &  &  &  &  &  &  &  &  &  &  &  &  &  & \ding{108}	 &  &  &  &  &  &  &  &  &  &  &  &  &  &  & 1/1 \\
Mc-29	 &  & \ding{108}	 & \ding{108}	 &  &  &  &  &  &  &  &  &  &  &  &  &  &  &  &  &  &  &  &  &  &  &  &  &  &  &  &  &  &  &  &  & 2/2 \\
\rowcolor{lightgray}Mc-38	 &  &  & \ding{108}	 & \ding{108}	 &  &  &  &  &  &  &  &  &  &  &  &  &  &  &  &  &  &  &  &  &  &  &  &  &  &  &  &  &  &  &  & 2/2 \\
\hline
T-3	 &  &  &  &  &  &  &  &  &  &  & {\uline\LEFTcircle}	 &  &  &  &  &  &  &  &  &  &  &  &  &  &  &  &  &  &  &  &  &  &  &  &  & 1/1 \\
\rowcolor{lightgray}T-7	 &  &  &  &  &  &  &  &  &  &  &  &  &  &  &  &  &  &  &  &  &  &  &  &  &  &  &  &  &  &  & \ding{108}	 & \ding{109}	 &  &  &  & 1/2 \\
T-19	 &  &  &  &  &  &  &  &  &  &  &  &  &  &  & \ding{109}	 &  &  &  &  &  &  &  &  &  &  &  & \ding{108}	 &  &  &  &  &  &  &  &  & 1/2 \\
\rowcolor{lightgray}T-26	 &  &  &  &  &  & 	 &  &  &  &  &  &  &  &  &  &  &  &  &  &  &  &  &  &  &  & \ding{108}	 &  &  &  &  &  &  &  &  &  & 1/1 \\
\hline
{\bf\# 1} & 1 & 6& 5 & 4 & 1 & 1 &0& 3 & 1 & 0 & 1 &0& 1 & 3 & 5& 3 & 0& 1 & 1 & 1 &6& 0 & 3 & 1 & 1 & 3 &11& 1 & 0 &0  &12 & 2 & 2 & 13 & 2 &  \\
{\bf\# 2} & 1 & 7& 10 & 6 & 1 & 1 &0& 4 & 1 & 0 & 14 &0& 15 & 12 & 32& 3 & 0& 1 & 1 & 1 &6& 7 & 4 & 2 & 2 & 3 &24& 2 & 0 &1  &43 & 19 & 6 & 25 & 4 &  \\
\bottomrule
		\end{tabular}
		{\scriptsize $\ast$ \ding{108} indicates that the bug is correctly fixed and \ding{109} indicates that the generated patch is plausible but not correct. \LEFTcircle means that the fix pattern can generate both correct patch and plausible patch for a bug. \uline{\ding{108}} and \uline{\ding{109}} denote that the bug can be partially fixed by the corresponding fix pattern. In the last column, we provide {\em x/y} numbers: {\em x} is the number of fix patterns that can generate correct patches for a bug, and {\em y} is the number of fix patterns that can generate plausible patches for a bug.
		{\bf\em Note that}, the bugs that can be plausible but incorrectly fixed by fix patterns are not shown in this table.
		{\bf \# 1}: number of bugs correctly fixed by a fix pattern.
		{\bf \# 2}: number of bugs plausible fixed by a fix pattern.
		}
	\end{threeparttable}
	}
\end{table*}

%% file: tables/exp2FixedBugs.tex
\begin{table*}[!t]
	\centering
	\scriptsize
	\setlength\tabcolsep{2pt}
	\caption{Comparing \toolname against the state-of-the-art APR tools.}
	\label{tab:comparison}
    	\begin{threeparttable}
			\begin{tabular}{l|c|c|c|c|c|c|c|c|c|c|c|c|c|c|c|c|c|c}
			\toprule
			\multirow{2}{*}{\bf Project} & \multirow{2}{*}{\bf jGenProg} & \multirow{2}{*}{\bf jKali} & \multirow{2}{*}{\bf jMutRepair} & \multirow{2}{*}{\bf HDRepair} & \multirow{2}{*}{\bf Nopol} & \multirow{2}{*}{\bf ACS} & \multirow{2}{*}{\bf ELIXIR} & \multirow{2}{*}{\bf JAID} & \multirow{2}{*}{\bf ssFix} & \multirow{2}{*}{\bf CapGen} & \multirow{2}{*}{\bf SketchFix} & \multirow{2}{*}{\bf FixMiner} & \multirow{2}{*}{\bf LSRepair} & \multirow{2}{*}{\bf SimFix} & \multirow{2}{*}{\bf kPAR} & \multirow{2}{*}{\bf AVATAR} & \multicolumn{2}{c}{\cellcolor{lightgray}{\bf TBar}} \\\cline{18-19}
			 & & & & & & & & & & & & & & & & & \cellcolor{lightgray}{Fully fixed}&\cellcolor{lightgray}{Partially fixed}\\
	        \hline
 Chart   & 0/7  & 0/6  & 1/4  & 0/2  & 1/6  & 2/2  & 4/7  & 2/4  & 3/7  & 4/4  & 6/8  & 5/8  & 3/8  & 4/8 & 3/10 &  5/12 & \cellcolor{lightgray}{\bf 9/14}& \cellcolor{lightgray}{0/4}\\
 Closure & 0/0  & 0/0  & 0/0  & 0/7  & 0/0  & 0/0  & 0/0  & 5/11  & 2/11  & 0/0  & 3/5  & 5/5  & 0/0  & 6/8 & 5/9 & 8/12& \cellcolor{lightgray}{\bf 8/12}&\cellcolor{lightgray}{1/5} \\
 Lang    & 0/0  & 0/0  & 0/1  & 2/6  & 3/7  & 3/4  & 8/12  & 1/8  & 5/12  & 5/5  & 3/4  & 2/3  & 8/14  & {\bf9}/13 & 1/8 &5/11 & \cellcolor{lightgray}{5/{\bf14}}& \cellcolor{lightgray}{0/3}\\
 Math    & 5/18  & 1/14  & 2/11  & 4/7  & 1/21  & 12/16  & 12/19  & 1/8  & 10/26  & 12/16  & 7/8  & 12/14  & 7/14  & 14/26 & 7/18 & 6/13 & \cellcolor{lightgray}{\bf 19/36}&\cellcolor{lightgray}{0/4} \\
 Mockito & 0/0  & 0/0  & 0/0  & 0/0  & 0/0  & 0/0  & 0/0  & 0/0  & 0/0  & 0/0  & 0/0  & 0/0  & 1/1  & 0/0 & 1/2 & {\bf2/2} & \cellcolor{lightgray}{1/2}& \cellcolor{lightgray}{0/0}\\
 Time    & 0/2  & 0/2  & 0/1  & 0/1  & 0/1  & 1/1  & {\bf2}/3  & 0/0  & 0/{\bf4}  & 0/0  & 0/1  & 1/1  & 0/0  & 1/1 & 1/2 & 1/3 & \cellcolor{lightgray}{1/3}& \cellcolor{lightgray}{1/2}\\
			\hline
		    Total &	5/27 & 1/22  & 3/17  & 6/23  & 5/35  & 18/23  & 26/41  & 9/31  & 20/60  & 21/25  & 19/26  & 25/31  & 19/37  & 34/56 & 18/49 & 27/53 & \cellcolor{lightgray}{\bf 43/81}&\cellcolor{lightgray}{2/{18}} \\\hline
P(\%)  & 18.5 & 4.5 & 17.6  & 26.1  & 14.3  & 78.3  & 63.4  & 29.0  & 33.3  & {\bf84.0} & 73.1  & 80.6  & 51.4  & 60.7 & 36.7 & 50.9 &\cellcolor{lightgray}{53.1} & \cellcolor{lightgray}{11.1} \\
		\bottomrule
		\end{tabular}
		{\scriptsize$^\ast$``{\bf P}'' is the probability of generated plausible patches to be correct. The data of other APR tools are excerpted from the corresponding work. {\bf kPAR}~\cite{liu2019you} is an open-source implementation of PAR~\cite{kim2013automatic}.
		}
		\end{threeparttable}

\end{table*}

%% file: tables/fpFixedBugs.tex
\begin{table*}[!h]
	\centering
	\scriptsize
	\caption{Per-pattern repair performance.}
	\resizebox{1\linewidth}{!}{
	\begin{threeparttable}
		\begin{tabular}{l|c|c|c|c|c|c|c|c|c|c|c|c|c|c|c|c|c|c|c|c|c|c|c|c|c|c|c|c|c|c|c|c|c|c|c}
			\toprule
			\multirow{2}{*}{} & \multirow{2}{*}{\rotatebox[origin=l]{90}{\bf FP1}} & \multicolumn{5}{c|}{\bf FP2} & \multirow{2}{*}{\rotatebox[origin=l]{90}{\bf FP3}} & \multicolumn{4}{c|}{\bf FP4} & \multirow{2}{*}{\rotatebox[origin=l]{90}{\bf FP5}} & \multicolumn{3}{c|}{\bf FP6} & \multicolumn{2}{c|}{\bf FP7} & \multicolumn{3}{c|}{\bf FP8} & \multicolumn{2}{c|}{\bf FP9} & \multicolumn{4}{c|}{\bf FP10}& \multicolumn{3}{c|}{\bf FP11} & \multirow{2}{*}{\rotatebox[origin=l]{90}{\bf FP12}} & \multicolumn{2}{c|}{\bf FP13} & \multirow{2}{*}{\rotatebox[origin=l]{90}{\bf FP14}} & \multicolumn{2}{c}{\bf FP15}\\\cline{3-7}\cline{9-12}\cline{14-30}\cline{32-33}\cline{35-36}
			 & & 1 & 2 & 3 & 4 & 5 & & 1 & 2 & 3 & 4 & & 1 & 2 & 3 & 1 & 2 & 1 & 2 & 3 & 1 & 2 & 1 & 2 & 3 & 4 & 1 & 2 & 3 & & 1 & 2 & & 1 & 2\\			\hline
{\bf Correct} & 1 & 4& 2 & 1 & 0 & 1 &0& 1 & 0 & 0 & 0 &0& 0 & 0 & 3 & 3 & 0& 0 & 0 & 1 &2& 0 & 1 & 1 & 1 & 1 &7 & 1 & 0 &0  &9  & 1 & 0 & 2 & 2 \\
{\bf Avg position*} & (1) & (16)& (1) &(5) & - & (5) & -& (5) & - & - & - &-& - & - & (23) & (16) & -& - & - & (9) &(1)& - & (2) & (62) & (6) & (1) & (12) & (18) & - &- &(5)  & (1) & - & (2) & (1) \\ \hline
{\bf Plausible (all)} & 1 & 7& 4 & 1 & 0 & 1 &0& 3 & 0 & 0 & 0 &0& 1 & 0 & 11& 4 & 0& 0 & 0 & 1 &4& 0 & 2 & 2 & 1 & 1 &12& 1 & 0 &0  &25 & 4 & 1 & 7 & 5 \\
{\bf Avg position*} & (1) & (12)$^\dagger$& (191) &(5) & - & (5) & -& (20) & - & - & - &-& (8) & - & (27)$^\dagger$ & (15) & -& - & - & (9) &(18)& - & (4) & (49) & (6) & (1) & (15)$^\dagger$ & (18) & - &- &(8)$^\dagger$  & (20) & (15) & (26) & (16)\\
			\bottomrule
		\end{tabular}
		{$^\ast$Average position of the exact buggy position in the list of suspicious statements yield by fault localization tool. {\bf $\dagger$} The exact buggy positions of some bugs cannot be yield by fault localizaiton tool.}
	\end{threeparttable}
	}
	\label{tab:fpFixedBugs}
\end{table*}

%% file: discussion.tex
\section{Discussion}
\label{sec:dic}
Overall, our investigations reveal that a large catalogue of fix patterns can help improve APR performance. However, at the same time, there are other challenges that must be dealt with: more accurate fault localization, effective search of relevant donor code, fix pattern prioritization. While we will work on some of these research directions in future work, we discuss in this section some threats to validity of the study and practical limitations of \toolname.

%

\subsection{Threats to Validity}
Threats to external validity include the target language of this study, i.e., Java.
Fix patterns studied in this paper only cover the fix patterns targeting at
Java program bugs released by the state-of-the-art pattern-based APR systems.
However, we believe that most fix patterns presented in this study could be
applied to other languages since fix patterns are illustrated as abstract syntax tree level.
Another threat to external validity could be the fix pattern diversity.
Our study may not consider all available fix patterns so far in the literature.
To reduce this threat, we systematically reviewed the research on
pattern-based program repair in the literature.
Nevertheless, we acknowledge that integrating more fix patterns may not necessarily lead to increased number of bugs that are correctly fixed. With too many fix patterns, the search space of fix patterns and patch candidates will explode. Eventually, the APR tool will produce a huge number of plausible patches, many of which might be validated before the correct ones~\cite{wen2018context}. A future research direction could be on the construction and curation of fix patterns database for APR.

Our strategy of fix pattern selection can be a threat to internal validity: it na\"ively matches patterns based on the AST context
around buggy locations. More advanced strategies would give a higher probability
to select appropriate patterns to fix more bugs.
Our approach to searching for donor code also carries some threats to validity:  \toolname focuses on the local buggy file, while previous works have shown that the adequate donor code, for some bugs, is available in other files~\cite{wen2018context,jiang2018shaping}. In future work, we will investigate the search of donor code beyond local files, while using heuristics to cope with the potential search space explosion.
Finally, the selected benchmark for evaluation constitutes another threat to external validity for assessment.
The performance achieved by \toolname on Defects4J may not be reached on a bigger, more diverse and more representative dataset. To address this threat, new benchmarks such as
Bugs.jar~\cite{saha2018bugs} and Bears~\cite{madeiral2019bears} should be investigated.

\subsection{Limitations}
\toolname selects fix patterns in a na\"ive way, it thus would be necessary to design a sophisticated strategy (such as bug symptom, bug type, or other information from bug reports) for fix pattern selection to reduce the noise from inappropriate fix patterns.
Searching donor code for synthesis patches is another limitation of \toolname, as the correct donor code for fixing some bugs is located in the code files that do not contain the bug~\cite{wen2018context, jiang2018shaping}.
If \toolname extends the donor code searching to other non-buggy code files, it will cause the search space explosion.

%% file: relatedwork.tex
\section{Related Work}
\label{relatedwork}
{\bf Fault Localization.} In general, most APR pipelines start with
fault localization (FL), as shown in Figure~\ref{fig:TBar}.
Once the buggy position is localized, ARP tools can mutate the buggy
code entity to generate patches.
To identify defect locations in a program, several automated FL techniques have been
proposed~\cite{wong2016survey}: slice-based~\cite{wong2010family,mao2014slice}, spectrum-based~\cite{abreu2009spectrum,perez2017test}, statistics-based~\cite{liblit2005scalable,
liu2006statistical}, etc.

Spectrum-based FL is widely adopted in APR systems since they identify bug
position at the statement level.
It relies on the ranking metrics (e.g., Trantula~\cite{jones2005empirical},
Ochiai~\cite{abreu2009practical}) 
to calculate the suspiciousness of each statement.
GZoltar~\cite{campos2012gzoltar} and Ochiai have been widely integrated into APR systems since their
effectiveness has been demonstrated in several empirical
studies~\cite{steimann2013threats,xie2013theoretical,xuan2014learning,pearson2017evaluating}.
As reported by Liu et al.~\cite{liu2019you} and studied in this paper,
this FL configuration still has a limitation on localizing bug positions.
Therefore, researchers tried to enhance FL techniques with new techniques, such as predicate switching~\cite{zhang2006locating,xiong2017precise} and test case purification~\cite{xuan2014test,jiang2018shaping}.

\noindent
{\bf Patch Generation.}
Another key process of APR pipelines is 
searching for another shape of a program (i.e., a patch)
in the space of all possible programs~\cite{le2012systematic,long2016analysis}.
If the search space is too small, it might not include the correct patches.~\cite{wen2018context}.
To reduce this threat, a straightforward strategy is to expand the search space, however, which could lead to other two problems:
(1) at worst, there still is no correct patch in it;
and (2) the expanded search space includes more plausible patches
that enlarge the possibility of generating plausible patches before correct ones~\cite{wen2018context,kui2018live}.

To improve repair performance, many APR systems have been explored to address
the search space problem.
Synthesis-based APR systems~\cite{long2015staged,xuan2017nopol,xiong2017precise}
explored to limit the search space on conditional bug fixes by
synthesizing new conditional expressions with variables identified from the buggy code.
Pattern-based APR tools~\cite{kim2013automatic,le2016history,saha2017elixir,long2017automatic,
durieux2017dynamic,le2017s3,liu2018mining,hua2018towards,jiang2018shaping,liu2019avatar}
are designed to purify the search space by following fix patterns to mutate buggy code entities with retrieved donor code.
Other APR pipelines focus on specific search methods for donor
code  or patch synthesizing strategies, to address the search space problem,
such as contract-based~\cite{wei2010automated,chen2017contract},
symbolic execution based~\cite{nguyen2013semfix},
learning based~\cite{long2016automatic,gupta2017deepfix,rolim2017learning,soto2018using,bhatia2018neuro,white2019sorting},
and donor code searching~\cite{mechtaev2015directfix, ke2015repairing} APR tools.
Various existing APR tools have achieved promising results on fixing real bugs, but there  is still an opportunity to improve the performance; for example, mining more fix patterns, improving pattern selection and donor code retrieving strategy, exploring a new strategy for patch generation, and prioritizing bug positions.

\noindent
{\bf Patch Correctness.}
The ultimate goal of APR systems is to automatically generate a correct patch
that can resolve the program defects.
In the beginning, patch correctness is evaluated by passing all test
cases~\cite{westley2009automatically,kim2013automatic,le2016history}.
However, these patches could be overfitting~\cite{qi2015analysis,le2018overfitting}
and even worse than the bug~\cite{smith2015cure}.
Since then, APR systems are evaluated with the precision of generating correct
patches~\cite{xiong2017precise,wen2018context,jiang2018shaping,liu2019avatar}.
Recently, researchers start to explore automated frameworks that can identify
patch correctness for APR systems automatically~\cite{xiong2018identifying, le2019reliability}.

%% file: conclusion.tex
\section{Conclusion}
\label{con}
Fix patterns have been studied in various scenarios to understand bug fixes in the wild. They are further implemented in different APR pipelines to generate patches automatically.
Although template-based APR tools have achieved promising results,
no extensive investigation on the effectiveness fix patterns was conducted.
We fill this gap in this work by revisiting the repair performance of fix patterns via a systematic study assessing the effectiveness of a variety of fix patterns summarized from the literature.
In particular, we build a straightforward template-based APR tool, \toolname, which we evaluate on the Defects4J benchmark.
On the one hand, assuming a perfect fault localization, \toolname fixes 74/101 bugs correctly/plausibly.
On the other hand, in a normal/practical APR pipeline, \toolname correctly fixes 43 bugs despite the noise of fault localization false positives. This constitutes a record performance in the literature on Java program repair. We expect \toolname to be established as the new baseline APR system, leading researchers to propose better techniques for substantial improvement of the state-of-the-art.